\documentclass[twocolumn]{aastex62}
\shorttitle{BoRG galaxies at z$\ge$7}
\shortauthors{Rojas-Ruiz et al.}

\usepackage{float}
\usepackage{natbib}
\usepackage{hyperref}
\usepackage{wasysym}
\usepackage{ragged2e}
\usepackage{graphicx}
\usepackage{units}
\newcommand{\sol}{$_{\odot}$}

\begin{document}


\title{\bf Probing the Bright End of the Rest-Frame Ultraviolet Luminosity Function at $z =$ 8--10 with \textbf{\emph{Hubble}} Pure-Parallel Imaging

}
\correspondingauthor{Sof\'ia Rojas-Ruiz}
\email{rojas@mpia.de}

\author{Sof\'ia Rojas-Ruiz}
\affiliation{The University of Texas at Austin, Department of Astronomy, \\
 2515 Speedway Boulevard Stop C1400, \\
Austin, TX 78712, USA}
\affiliation{Max-Planck-Institut f\"ur Astronomie, K\"onigstuhl 17, D-69117, Heidelberg, Germany}

\author{Steven L. Finkelstein}
\affiliation{The University of Texas at Austin, Department of Astronomy, \\
 2515 Speedway Boulevard Stop C1400, \\
Austin, TX 78712, USA}
\author{Micaela B. Bagley}
\affiliation{The University of Texas at Austin, Department of Astronomy, \\
 2515 Speedway Boulevard Stop C1400, \\
Austin, TX 78712, USA}
\author{Matthew Stevans}
\affiliation{The University of Texas at Austin, Department of Astronomy, \\
 2515 Speedway Boulevard Stop C1400, \\
Austin, TX 78712, USA}
\author{Keely D. Finkelstein}
\affiliation{The University of Texas at Austin, Department of Astronomy, \\
 2515 Speedway Boulevard Stop C1400, \\
Austin, TX 78712, USA}
\author{Rebecca Larson}
\affiliation{The University of Texas at Austin, Department of Astronomy, \\
 2515 Speedway Boulevard Stop C1400, \\
Austin, TX 78712, USA}
\author{Mira Mechtley}
\affiliation{School of Earth and Space Exploration, Arizona State University, P.O. Box 871404,  \\
 Tempe, AZ 85287-1404, USA}
 \author{James Diekmann}
\affiliation{The University of Texas at Austin, Department of Astronomy, \\
 2515 Speedway Boulevard Stop C1400, \\
Austin, TX 78712, USA}
%


\begin{abstract}
Looking for bright galaxies born in the early universe is fundamental to investigating the Epoch of Reionization, the era when the first stars and galaxies ionized the intergalactic medium. 
We utilize {\it Hubble Space Telescope} pure parallel imaging to select galaxy candidates at a time 500 to 650 million years after the Big Bang, which corresponds to redshifts $z \sim8-10$. These data come from the Brightest of Reionizing Galaxies Survey (BoRG) Cycle 22 dataset, which consists of pure-parallel imaging in $\sim90$ different lines of sight that sum up to an area of $\sim420$ arcmin$^2$. This survey uses five filters and has the advantage (compared to the Cycle 21 BoRG program) of including imaging in the $JH_{140}$ band, covering continuous wavelengths from the visible to near-infrared ($\lambda$ = 0.35$\mu$m -- 1.7$\mu$m).  This allows us to perform reliable selection of galaxies at $z \ge8$ using the photometric redshift technique. We use these galaxy candidates to constrain the bright end of the rest-frame ultraviolet luminosity function in this epoch.  These candidates are excellent targets for follow-up observations, particularly with 
the {\it James Webb Space Telescope}.
\end{abstract}

\vspace{0.5cm}


\section{Introduction}\label{intro}
Discoveries in the high-redshift universe have been possible within the past decade by using the Wide Field Camera 3 (WFC3) on the {\it Hubble Space Telescope} to detect galaxies born only $\sim$0.5--1 Gyr after the Big Bang. Deep surveys such as the Cosmic Assembly Near-infrared Deep Extragalactic Legacy Survey (CANDELS; \citealt{Grogin2011}, \citealt{koekemoer2011}), Hubble Ultra Deep Field (HUDF; \citealt{Oesch2010b}, \citealt{Bouwens2010a}, \citealt{ellis13}),  
Hubble Frontier Fields (HFF; \citealt{Lotz2017}), 
and the Hubble Infrared Pure Parallel Imaging Extragalactic Survey (HIPPIES; \citealt{Yan2011}) 
have been crucial for identifying galaxies at redshifts as high as $z \sim11$ \citep{Oesch2016}. 
Finding these distant galaxies and quantifying their abundance places crucial constraints on the conditions of ionizing sources during the epoch of reionization.

At about 380,000 years after the Big Bang, the universe had cooled down enough to form neutral hydrogen, which is capable of absorbing high-energy photons and thus making the universe opaque. Later, the first stars and galaxies radiated high-energy photons in sufficient amounts to escape the galaxy and ionize the surrounding neutral hydrogen. Thus, creating ionized bubbles around the galaxies for light to travel through to the intergalactic medium (IGM) and illuminate the universe we observe today.  Previous studies of quasar absorption spectra have shown that this process of reionization ends by $z \sim6$ \citep{McGreer2015}. However, when reionization begins and how fast this process happens as the universe evolves is still not well-constrained. 

The prevailing theory for reionization is one where it starts slowly, with the bulk of IGM ionization taking place at $z <$ 8, and ends rapidly by $z \sim6$ \citep[e.g.,][]{Robertson2015,Finkelstein2015,bouwens15b,Mason2019}. This model assumes that all galaxies have a comparable escape fraction, typically assumed to be 10-20\% \citep[e.g.,][]{Finkelstein2012b,Robertson2015}.  However, escape fractions this high are not observed for the bulk of bright galaxies at $z <$ 3, with deep imaging studies finding upper limits on the average escape fraction of $<$5\% \citep[e.g.][]{grazian17}, and spectroscopic stacking work finding no higher than $\sim9$\% \citep{steidel18}.  \citet{finkelstein19} tried to reconcile this discrepancy by proposing a model of reionization where the faintest galaxies have high escape fractions \citep[motivated by simulations, e.g.,][]{paardekooper15,xu16}, and bright galaxies contribute little to reionization.  This model results in a reionization process which starts earlier, and proceeds more smoothly than the prevailing model. These scenarios differ the most at $z \sim9$ where this model predicts reionization should be $\sim$50\% complete, while previous work would predict $\sim$20\%.  Placing robust constraints on the $z \sim9$ galaxy population is the first step towards understanding whether reionization was truly well underway by that time.  These observations also place key constraints on models of galaxy growth at early times \citep[e.g.,][]{yung19,vogelsberger19}.

A number of studies to date have photometrically selected galaxies at $z >$ 9, yet with discrepant results.  Some studies find a comparable number of galaxies at $z \sim$ 9--10 as would be expected from a simple extrapolation of the $z =$ 4--8 luminosity function \citep[LF; e.g.,][]{coe13,McLeod2015}, while others find fewer than expected, concluding that there is a sharp downturn in detectable star-formation activity at $z >$ 8 \citep[e.g.,][]{Bouwens2016,Oesch2018}.  These differences could be due to a number of effects, including the difference in datasets, and varying galaxy selection and analysis methods.  Cosmic variance may also play a role, as these studies are typically relegated to a few contiguous fields on the sky.

The Brightest of Reionizing Galaxies Survey (BoRG; \citealt{Trenti2011}) is a {\it Hubble Space Telescope} ({\it HST}) program aimed to improve constraints on the population of galaxies at $z >$ 8 by randomly sampling the sky. This survey consists of imaging data using {\it HST's} pure-parallel mode, observing with WFC3 while another {\it HST} instrument (typically the Cosmic Origins Spectrograph) is observing a nearby primary target. This method randomly samples the night sky, reducing systematic uncertainties due to cosmic variance and improving the fidelity of population estimates for high-redshift galaxies \citep{Trenti2012}. 

In this study we make use of the Cycle 22 BoRG program, known as BoRG[z910], which includes filter coverage that allows for more robust selection of $z >$ 9 galaxies. Previous analyses of BoRG[z910] reported in \citet{Calvi2016} and \citet{Morishita2018} both find high-redshift galaxies using first a Lyman break color selection, and then apply a photometric redshift measurement. However, the color selection box might exclude potential candidates and therefore in this study we instead rely exclusively on the photometric redshift technique aiming to produce a more complete sample of high-redshift galaxy candidates. We describe our data reduction and photometry in \S 2, and our photometric redshift measurements in \S 3.  In \S 4, we describe our galaxy selection process, and summarize our sample in \S 5, where we also compare our results to previous work.  Our luminosity function is presented in \S 6, and our conclusions are given in \S 7.
Throughout this work we use the cosmological parameters H$_0$= 70.2, $\Omega_M$= 0.275, and $\Omega_\Lambda$ = 0.725.

\section{Data set}\label{dataset}
\begin{figure}[t!]
\epsscale{1.3}
\plotone{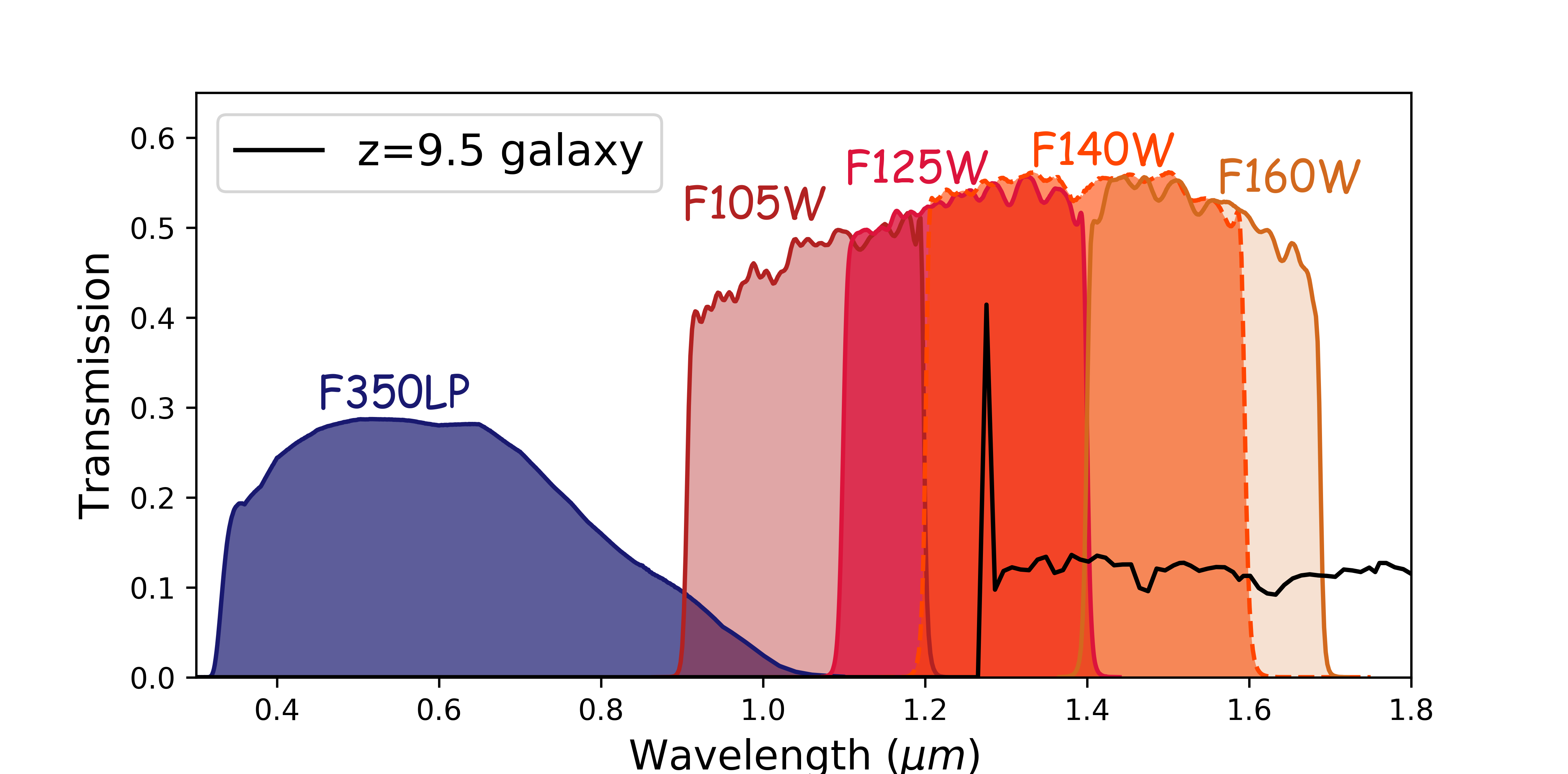}
\caption{Transmission curves of the filters in the WFC3/UVIS (F350LP) and WFC3/IR (F105W, F125W, F140W, F160W) cameras from the BoRG[z910] Cycle 22 dataset. This set of filters has the advantage of covering continuous wavelengths from 0.35--1.7 $\mu$m, making the photometric redshift technique a  reliable method for selecting Lyman-break galaxy candidates at z $>$ 7. The spectra shown in black corresponds to a galaxy model at z $=$ 9.5, which is similar to the galaxies we search for in this study.  The filter F140W, not available in the Cycle 21 BoRG survey, allows for a more precise measurement of the wavelength at which the Lyman break is observed, giving a better fit to the galaxy spectral energy distribution (SED) and thus finding a more accurate redshift probability for galaxies at $z \sim$ 9--10.}\label{trans}
\end{figure}

The BoRG survey has been implemented in multiple forms. The first part of the survey, from Cycle 21, was optimized for finding galaxies at $z \sim8$ using the wide-band filters F606W, F098M, F125W, and F160W \citep{Trenti2011,Bradley2012}. The Cycle 22 survey BoRG[z910], which we use here, adds the F140W filter to improve the selection of galaxies at $z >$ 9 (See Figure \ref{trans}). 
This survey uses both WFC3 cameras (IR and UVIS) to observe each pointing in the wide-band filters F350LP (UVIS), F105W, F125W, F140W, and F160W (IR), covering a continuous wavelength range from the optical to near-infrared at $\lambda$ = 0.35$\mu$m - 1.7$\mu$m (Figure \ref{trans}).  This modest amount of photometric information allows us to utilize the photometric redshift technique to select galaxy candidates at $z \sim$8-10, which is discussed in Section \ref{methods}.

These pure-parallel observations are taken randomly and at different points of the sky, reducing biases from cosmic variance and giving better statistics on the population of high-redshift galaxies at different redshifts ($z \sim$ 7--11). One unfortunate byproduct is that the data are not dithered, which presents analysis challenges due to the abundant hot pixels present in the data.  

We downloaded all data from the Cycle 22 BoRG program from the {\it HST} MAST/HLSP archive\footnote{DOI:\dataset[10.17909/T9QC7K]{\doi{10.17909/T9QC7K}}}, which consisted of five-band imaging in 92 fields. The field ``par1127+2652" (we will denote fields in this notation, ``par" denoting parallel, followed by four digits denoting the right ascension hours and minutes, and four digits [with a sign] denoting the declination degrees and minutes) has a guiding star acquisition failure as reported in the first analysis of this survey by \cite{Calvi2016} and is thus not included in this analysis. Our reduction combined two overlapping BoRG fields, therefore leaving a total of 90 fields to analyze. Refer to Table \ref{fields} for complete information of all 90 fields analyzed from BoRG[z910] in this study. 
More information about the Borg[z910] data are available at the survey website.\footnote{\url{http://borg.astro.ucla.edu/about/}} 

\subsection{Data Reduction}

We obtained the data from Space Telescope Science Institute (STScI) {\it HST} MAST archive in the form of flt images, which have already been corrected for instrumental response (e.g., flat-fielded). To further reduce these data, we used a reduction pipeline (by M.\ Mechtley) built for HIPPIES program in \cite{Yan2011}. This pipeline is thus custom-built for pure-parallel observations such as BoRG, to generate stacked science images, weight and RMS maps.  
After identifying which files belong to the same fields, 
this pipeline drizzles the images to a common pixel grid and runs the software ``Source Extractor'' \citep{Bertin1996} to make a catalog of pixel positions of sources in each exposure. An MCMC sampler then calculates the pixel shifts between the Source Extractor catalogs which are saved into a Multidrizzle shift file to combine the individual exposures using the calculated shifts, with an output pixel scale set to 0\farcs1.

Finally, the pipeline creates RMS maps to calculate uncorrelated background noise in the images. 
This step is accomplished by masking out the objects in an image, measuring the
auto-correlation of several sections of blank sky, and using the average of these to derive the proper scaling between the weight map and the true inverse variance of the background pixels.
The program then scales the weight map by this amount and takes one over its square root to derive an RMS map, where pixel values of zero weight have their RMS value set to 10000. A good measurement of the weight map becomes important in the visual inspection analysis of galaxy candidates from this survey, which is further explained in Section \ref{criteria}.


\subsection{Source Extractor}
We use the Source Extractor software tool to measure photometry of sources in this work.  Source Extractor can measure the flux and flux error from the sources in an image in different aperture sizes and shapes. We follow previous high-redshift studies \citep[e.g.,][]{Bouwens2010a,Finkelstein2010} by measuring object colors in a small Kron elliptical aperture (PHOT\_AUTOPARAMS of 1.2, 1.7), measuring the total flux in the F160W image in the default large (2.5, 3.5) Kron (or AUTO) aperture, which is tuned to measure the total flux from a source, at the cost of increased noise from the larger number of pixels in the aperture. 
We also measure the flux in a circular aperture of 0$\farcs$4 in diameter, which will not contain the total flux of the source, but serves as a high signal-to-noise measurement of the significance of flux present at a given wavelength.  This is relevant for our methods of selecting candidate  high-redshift galaxies described in Section \ref{methods}. Importantly, we can compare the ratio between the AUTO aperture sizes and the 0$\farcs$4 circular aperture to help identify point-like sources such as stars or bad pixels. 

To run Source Extractor it is necessary to set a detection image, which in this case is the F160W image (hereafter referred to as $H_{160}$) as we expect to detect our galaxies of interest at the highest signal-to-noise in this filter, as it is fully redward of the Ly$\alpha$ break for our full redshift range of interest ($z \lesssim$ 10.6). 
Source Extractor identifies sources in this image, creates a segmentation map in order to assign pixels to sources, and then creates a catalog with RA/DEC and pixel position, flux and flux errors in all given apertures of the sources in the images.  We cycle through all five filters as the measurement images.  We perform an additional Source Extractor run with the default Kron aperture setting to calculate aperture corrections,  derived as the ratio between the fluxes from the large to small Kron apertures (derived in the $H_{160}$, and applied to all filters).  
After creating the catalogs we correct for Galactic dust extinction in all filters using the attenuation curve described in \cite{Cardelli1989}. To calculate the E(B-V) due to galactic extinction we use the values from \cite{Schlafly2011}, obtained via the IRSA web tool.\footnote{\url{http://irsa.ipac.caltech.edu/applications/DUST}}
We apply these corrections to create a new catalog that will be used to calculate photometric redshifts. We clean the catalog of sources appearing near the edges, as well as objects with negative aperture corrections and objects with only negative fluxes, both typically indicating bright nearby objects.

\subsection{Noise Calculation}\label{noisecalc}
Due to the pure-parallel nature of this survey, the data are not dithered by the telescope. \cite{Calvi2016} presented an argument on dithered versus undithered data specifying that the quality was similar in both cases. Nevertheless, the undithered data presents problems in the noise calculation of the images, as many bad pixels cannot be removed (see Section \ref{criteria}). Additionally, the flux errors calculated with Source Extractor may not be reliable.
Therefore, we decided to directly calculate more reliable noise estimates for all objects in the 90 fields we analyze.

\begin{figure}[t!]
\centering
 \includegraphics[width=0.45\textwidth]{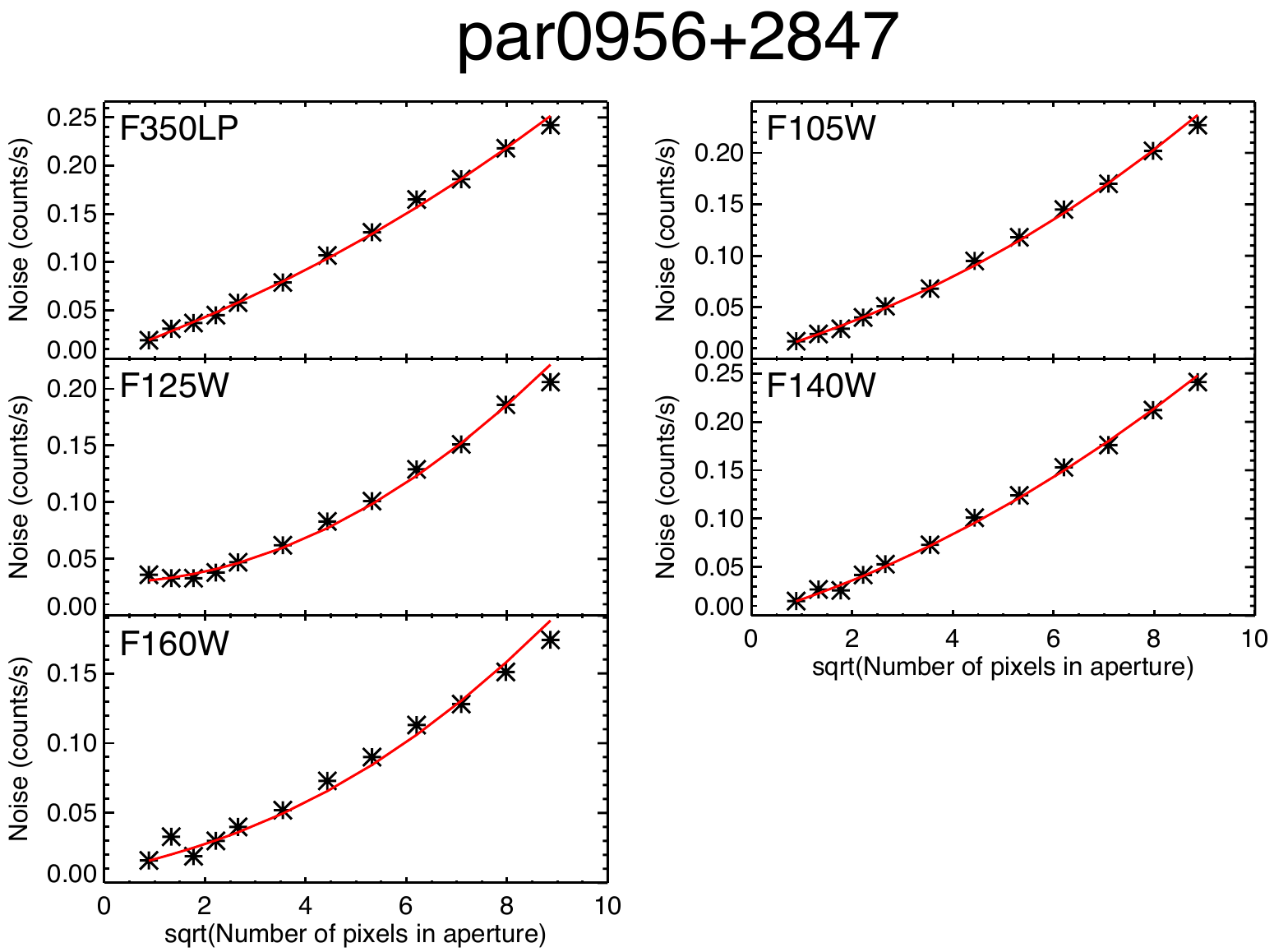}

\caption{Example of noise calculated for field par0956+2847 in the five images corresponding to each filter using the noise equation in Equation \ref{noise_eq}. The number of pixels in a given aperture varies depending on the size and each area gives a different estimate for the calculated noise. As expected, in a larger aperture size, the noise is higher. From the interpolated data points, the best fit correlating noise and aperture size is shown with the red line and was found using the free parameters $\alpha$ and $\beta$. This calculation is essential for correcting the flux errors measured in the image since these depend on aperture size. }\label{error_plot}
\end{figure}

We calculate the noise by randomly placing circular apertures of different sizes across each image, avoiding real sources using the segmentation map to identify empty portions of the image. We then use the method described in \cite{Papovich2016} to empirically measure how the image noise depends on the number of pixels in the aperture. This relation is based on the pixel aperture size N, a pixel-to-pixel standard deviation $\sigma_1$ and four free parameters $\alpha$, $\beta$, $\gamma$, $\delta$, which describe how the noise increases with number of pixels in the aperture. However, when trying to find the best fit to the correlation between aperture size and measured noise (Figure \ref{error_plot}), we found we could obtain a good fit using only two of the free parameters $\alpha$ and $\beta$. The equation that best describes the noise distribution in the BoRG fields is: 
\begin{equation}
\sigma_n = \sigma_1(\alpha N^\beta) 
\label{noise_eq}
\end{equation}
We measure $\sigma_1$, $\alpha$, and $\beta$ for all five filters in every field, and use these values to calculate the noise in any aperture of interest (using the semi-major and semi-minor axis calculated from the Source Extractor catalog for the Kron apertures). 

\section{Redshift Acquisition Method}\label{methods}
Galaxies at high redshift are challenging to find because of their distance. However, it is possible to observe the signature from bright star-forming galaxies by looking for the Lyman-break.
This occurs because photons with smaller wavelengths than the Lyman limit will be absorbed by neutral gas in the galaxy, or along the line-of-sight. Due to strong ``Gunn-Peterson" absorption, the intergalactic medium can also absorb photons with wavelengths ranging from  $\lambda_{rest} $ = 912\AA  $\ $ -  1216\AA\ \citep{Gunn1965}. Therefore, at $z \gtrsim$ 5, the Lyman break we observe is at $\lambda_{rest} $ = 1216\AA, which is traditionally referred to as the Ly$\alpha$ break. One can use this break to find galaxies at high redshift via color selection \citep{Steidel1996}, where the galaxies are found by identifying a region in a color-color space which is inhabited by galaxies at the method of interest (usually a red color in a pair of filters bracketing the Lyman/Ly$\alpha$ break, and a blue color in a pair of filters just red-ward of the break). A related method for finding these breaks involves fitting a suite of model template spectra to the observed spectral energy distribution (SED) of the galaxy candidate,  finding the redshift for which the model fits best, known as photometric redshift fitting.  The first analyses of BoRG[z910] reported in \citet{Calvi2016} and \citet{Morishita2018} both rely on first doing a Lyman break color selection, and then a photometric redshift measurement for identified candidates.

To select our own galaxy candidates in the BoRG[z910] survey, we choose to rely solely on the photometric-redshift technique to select high-redshift galaxies, in order to provide a more inclusive and potentially more complete selection, avoiding the potential exclusion of sources which may have scattered just outside the color-color selection box. The downside of either method is that, with these noisy pure parallel data we obtain a very long list of galaxy candidates that pass our selection criteria, where most are unlikely to be true high-redshift galaxies. We make use of machine learning, described in Section \ref{machine_learn}, to avoid human bias when cleaning the catalog of these likely interlopers.\\

\subsection{EAZY}
We measured photometric redshifts using the ``Easy and Accurate Z$_{phot}$ from Yale" (EAZY; \citealt{Brammer2008}) software, which calculates photometric redshifts based on different SED models of known galaxy types. The catalog used for running EAZY includes flux and flux uncertainty values for each source in all five filters. EAZY calculates a redshift probability distribution function $P(z)$ using the measured $\chi^2$ between the observed photometry and a given model. The template set used (known as EAZY\_v1.1\_lines) includes empirical SED templates from \citep{Fioc1997}, a dust-dominated galaxy \citep{Maraston2005}, and a high equivalent width nebular emission line galaxy \citep{Erb2010}.

We fit all SED models simultaneously (allowing combinations of templates).  EAZY has the ability to use a luminosity prior, which are derived from semi-analytic models of typical galaxies at a specific redshift. These are used to avoid biased selection of low-redshift galaxies at higher redshifts of $z \sim$ 3--6. However, the study of galaxy properties at  $z >$ 6 is constrained by the lack of observations, and \cite{Salmon2018} provide an argument of how priors are not well-studied for higher redshift galaxies, and when added tend to give lower-redshift solutions. Therefore we prefer to assume a flat luminosity prior. Here we note that \cite{Morishita2018} used a prior in EAZY for their galaxy selections in BoRG[z910]. The impact of including or not including a prior for finding galaxies can be better evaluated in the future and ideally with spectroscopically confirmed galaxies at high-redshift from different studies. However, it is important to note that we do select one of the galaxies in \cite{Morishita2018} as a high-redshift candidate, which we discuss further below.

\section{Selection Criteria} \label{criteria}
Using the results from EAZY, we developed a set of selection criteria to construct our final sample of $z =$ 7--11 galaxy candidates.  While our primary interest is $z >$ 9, this filter set can effectively select galaxies at $7 < z <$ 9, therefore we make use of this wider redshift range. The criteria we require for galaxy candidate selection relies on both the EAZY output and the galaxy photometry measured in the Source Extractor output catalogs.

We require that the integrated $P(z)$ from $7 < z <$ 11 be $\geqslant$ 0.6, thus dominating more than half the integral of the redshift probability distribution, strongly implying $z \geq$ 7. We also add the constraint that the integral of the primary $P(z)$ peak should be more than 50\% of the total integrated $P(z)$.
To ensure that a candidate is a robustly detected real source, we imposed as a selection criteria signal-to-noise (S/N) thresholds in the F140W and F160W images.  As we are using these criteria to establish the validity of a source as a real object, we use the 0$\farcs$4 aperture fluxes and errors for this measurement. Sources must satisfy $S/N>$ 5 in at least one of these filters, and $S/N>$ 3.5 in both filters.
We elected not to initially restrict the S/N in the F350LP band, as significant F350LP flux should have resulted in EAZY preferring a lower redshift (though we note that we eventually do require a non-detection here; see below).  Finally, we restricted our sample to $H >$ 22, eliminating bright stellar interlopers, while still accounting for expected magnitudes of high-redshift galaxies. Here we also note that the brightest candidate presented in Table \ref{tab_candidates} is more than three magnitudes fainter than a possible lensed high-redshift galaxy with mag$_{H_{160}} =$ 22, so it is very unlikely that we would find this in the data.

From this set of selection criteria, we have an initial sample of $\sim$600 sources.  Upon inspection most of these initial candidates were clearly bad pixels, diffraction spikes, or sources appearing in all five filters. 
While we could use this visual inspection to remove these spurious or otherwise improperly-measured sources, visual inspection is not a perfectly reproducible process for making decisions about galaxy candidates. Therefore, we decided to explore a simple machine learning algorithm to clean our sample of these spurious sources.\\


\begin{figure*}[ht!]
\centering
  \includegraphics[width=1\textwidth]{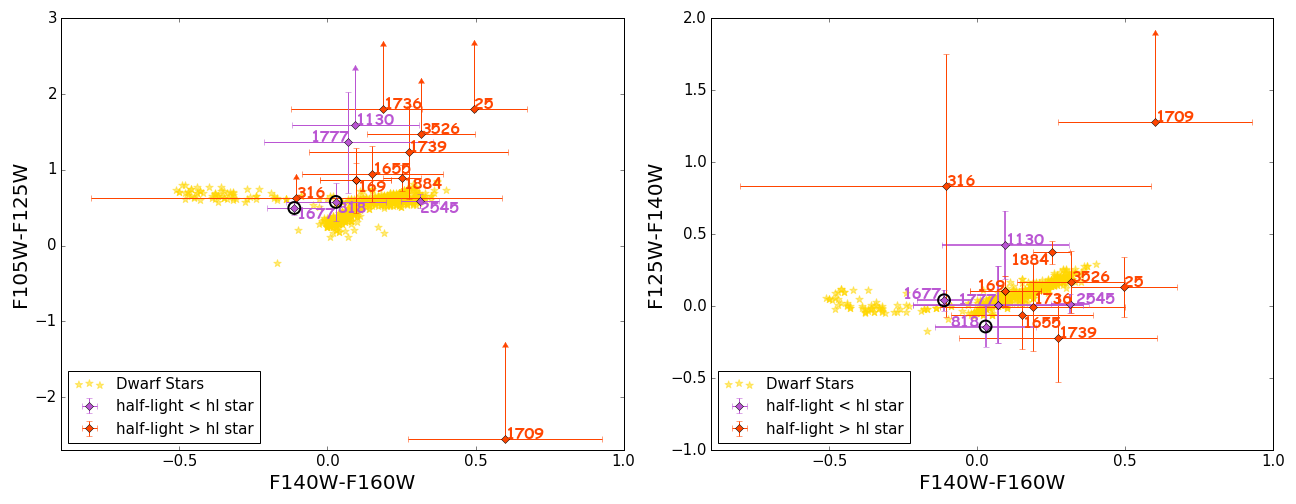}
\caption{{\it Left:} F105W-F125W vs. F140W-F160W color-color plot, used to diagnose stellar contamination. Known dwarf stars are shown in yellow (from the IRTF SpeX Library), with other symbols denoting our galaxy candidates after machine learning and visual inspection cuts. Candidates shown with orange symbols are well-resolved as they have a half-light radius bigger than that of a typical star (denoted as hl star) in this survey and are kept for the final catalog regardless of color. Candidates shown in purple have a smaller half-light radius so they are not well-resolved but passed our selection criteria. {\it Right:} Similar to the left panel,  showing F125W-F140W vs. F140W-F160W. Candidates with ID's of 1677 and 818, shown with black circles, are unresolved and have colors similar to dwarf stars in both panels.  These are thus removed from the final catalog due to their high probability of being stellar contaminants. While candidate 1130 does not appear well-resolved, it also does not have colors consistent with those of a star, therefore we keep it in our final sample. Candidates 1777 and 2545 have colors $>$1$\sigma$  different than the colors of dwarf stars in at least one of these plots, therefore we also include these candidates in the final sample. }\label{star_yj}
\end{figure*}

\subsection{Machine Learning}\label{machine_learn}
We used our initial round of visual inspection to categorize sources as  ``good" or ``bad" high-redshift galaxies. We then fed these classifications into the Python machine-learning algorithm ``DecisionTreeClassifier.'' We divide the machine learning process into two stages: In the first, we utilized the observables S/N$_{F350LP}$, half-light radius, and stellarity. The algorithm produced good results when restricting to sources with half-light radius 1.0 $\leq$ r  $<4.0$ pixels, which recovered all good sources analyzed by eye, reducing the galaxy candidate catalog by 86\% (from 579 to 82 candidates). Stellarity and S/N$_{F350LP}$ were not conclusively useful from the machine learning algorithm.

The second stage included an additional cut on the sources recovered from stage one. In this case we decided to use a S/N$_{350}$  $< 2.0$ cut because this is the  drop-out filter and therefore should not have any significant signal for the galaxies at z$>$7 we are aiming to find. 
We also increased our brightness cut to $H >$ 24, as many of the 22 $< H <$ 24 were due to bad pixels, or due to large galaxies more likely to be at lower redshift (indeed, these had significant $z \sim$ 2 peaks in the $P(z)$).  Furthermore, some of the sources remaining after stage one were clearly bad even though they had a half-light radius r $\gtrsim$ 1.0 pixel. From a last visual analysis, we found that we could remove the bulk of the obviously spurious sources by setting the half-light radius cut to $r_{0.5}>$ 1.2 pixels. While this 1.2 pixel cut could in principle remove real extremely compact galaxies from our sample, as shown by \citet{kawamata2018}, bright galaxies should be larger than this in size. 
After applying these restrictions for selecting candidates, we were left with 56 sources, which we visually analyzed again. The majority of sources in this catalog were still spurious coming from hot pixels in the weight map, or were parts of diffraction spikes from nearby stars. We also found that the SED of some sources definitely looked more similar to a star since the flux peaked in the $JH_{140}$ with a blue $JH_{140}$-$H_{160}$ color. After this final round of cleaning, we were left with a shorter catalog of 14 high-redshift galaxy candidates. 

\subsection{Removing Stars} \label{stars}
The wavelength coverage of the filters in BoRG are also sensitive to dwarf stars with the spectral classification of M, L, and T. The most evident way of identifying stars in BoRG[z910] with our methodology is by looking at the SED of the source and identifying  blackbody-like emission. For instance, when there is no detection in the first two filters, then there is an apparent Lyman break in the $J_{125}$ band, a peak of flux in the $JH_{140}$ band, but the flux drops back down in the $H_{160}$ band. This aspect was not easily evaluated from the machine learning algorithm or the visual inspection because the uncertainties in the flux measurements could also indicate that the source was a possible galaxy. In order to find a more quantitative method of removing these interlopers, we checked if the source is resolved through the measurement of the half-light radius, by comparing the half-light radii of the sources to that of the stars in the data. 

To measure the half-light radii of stars in the $H$-band, we ignored the most over-crowded fields as it was harder to identify the individual stars. From the remaining fields selected, we find the resolved stars from a magnitude versus half-light radius analysis of the sources in the field, this way avoiding the selection of galaxies or bad pixel sources. From this analysis we select the half-light and magnitude range for stars and calculate the median half-light radius which is found to be $r_{0.5}$ = 1.39 pixels, with a standard deviation $\sigma_{r_{0.5}}$ = 0.13 pixels.

The galaxy candidates with a measured half-light radius comparable to (or smaller than) that of the typical star in the survey are not resolved sources, and thus need to be more carefully analyzed to see if they have colors similar to stars.  We use the IRTF SpeX Library of MLT dwarf stars developed by \cite{SpexLibrary} to place MLT dwarfs on a color-color plot alongside our sources, as shown in Figure \ref{star_yj}. For the candidates with smaller $r_{0.5}$ than a star and which have colors similar to those of dwarf stars within 1$\sigma$ (e.g., co-located on both of these plots), we conclude that they are likely a star, and remove them from our catalog. From this analysis our candidates ID$=$1677 and ID$=$818 appear to be likely stars, and are thus removed from our sample.  All other candidates are not considered stars with this test or by their SEDs as shown in Figures \ref{cand1} and \ref{cand2}.

\subsection{{\it Spitzer}/IRAC Photometry}\label{irac}
As a final screen against contaminants in our sample, we include available \textit{Spitzer}/IRAC 3.6$\micron$ imaging to our 
analysis of high redshift galaxy candidates.
Passively-evolving and very dusty galaxies at $z \sim2-3$ can both exhibit 
similar $J-H$ colors as galaxies at $z\gtrsim8$ and be undetected in bluer
filters at the depth of the BoRG survey. As these lower-redshift galaxies 
are very red, we can expect them to be relatively bright at longer 
wavelengths. The addition of even shallow imaging at 3.6$\micron$ can 
therefore help distinguish between truly high redshift galaxies and 
lower-redshift contaminants. Five of the BoRG[z910] fields that contain our
high-redshift candidates have been observed with IRAC. In this section, we 
discuss the measurement of IRAC photometry for candidates in these fields.

\begin{figure*}[ht!]
\epsscale{1.0}
\plotone{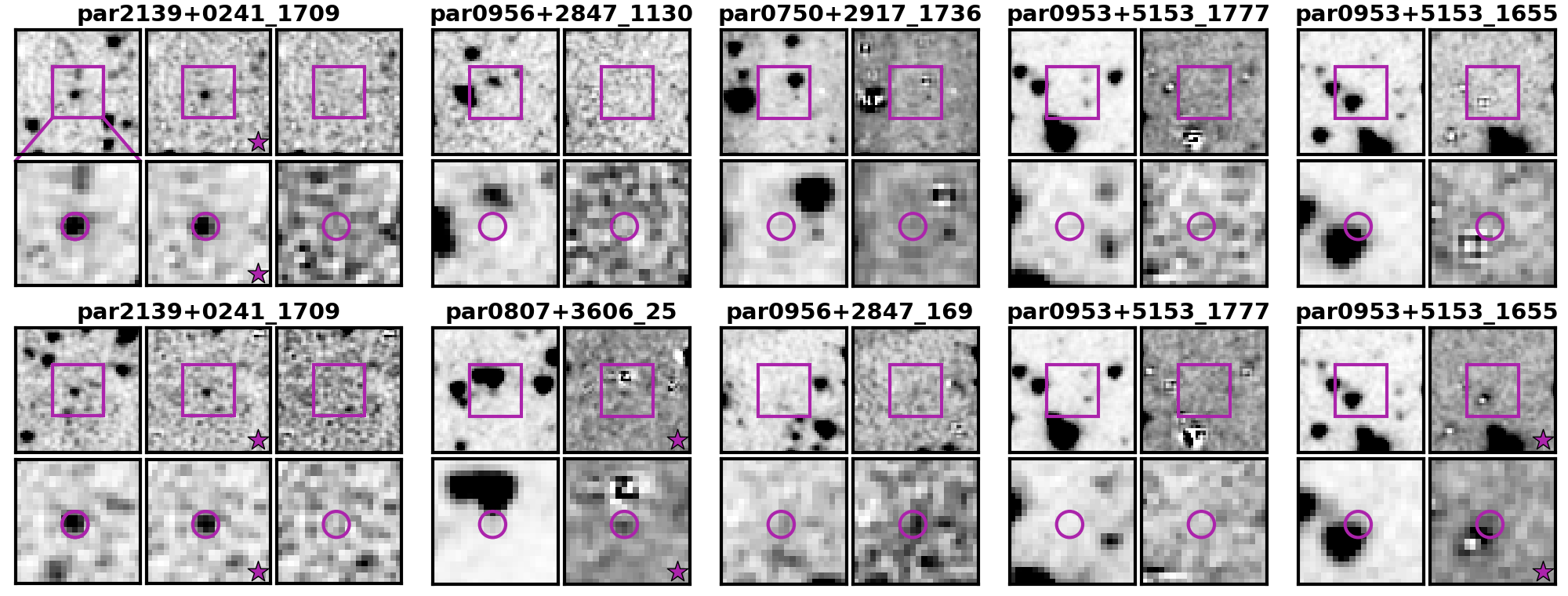}
\caption{The \textit{Spitzer}/IRAC 3.6\micron\ image stamps and residual maps. For each candidate, we show two rows of stamps and residual maps: $31\farcs6$ on a side in the top row and $12\farcs6$ in the bottom row. In each set of panels, the left column displays the 3.6\micron\ image that we modeled with \textsc{Galfit}, using a PSF constructed from unresolved bright sources in these five fields. The residual maps are shown in the right column(s). The $12\farcs6$ stamps zoom in to highlight the residual map at the position of the candidate. The purple stars indicate observations for which \textsc{Galfit} attributed some of the IRAC light to the candidate's position (though this flux was not necessarily above the noise). In these cases, the model of the candidate is not subtracted from the residual map. In all cases, we measure photometry on the residual maps (with the candidate left in) in circular apertures 4.6 pixels in diameter ($1.4\times$ the FWHM of the IRAC $3.6\mu$m point response function), indicated by the purple circles. For candidates observed multiple times with IRAC, we include each observation here as a separate set of panels.\label{iracfig}}
\end{figure*}

The \textit{Spitzer}/IRAC imaging was obtained by three programs, each 
designed to follow-up high-redshift candidates identified by other teams in 
the BoRG fields. Observations of par0807+3606, par0953+5153, par0956+2847,
and par2139+0241 were obtained as part of programs 12058 and 14130
(PI: R. Bouwens), and par0750+2917 was observed as part of program 
14233 (PI: T. Morishita). Between these programs, two fields 
(par2139+0241 and par0953+5153) were observed twice at different position 
angles, which we include in our analysis as independent measurements. 
We downloaded the Level 2 (PBCD) mosaic images for each field from the 
Spitzer Heritage Archive hosted by \textsc{IRSA\footnote{\url{https://sha.ipac.caltech.edu/applications/Spitzer/SHA/}}}.
The mosaic images are on a $0\farcs6$ pixel scale. The typical exposure time per field is $1-2$ hours, and we have measured the background noise empirically as described below, finding an average $5\sigma$ depth of 23.6 mag.

More than half of the high-redshift candidates are in crowded regions with 
one or multiple nearby neighbors, complicating the photometry in the 
lower-resolution IRAC images. We therefore use the two-dimensional 
image-fitting software \textsc{Galfit\footnote{\url{https://users.obs.carnegiescience.edu/peng/work/galfit/galfit.html}}}
\citep[v3.0;][]{peng2010} 
to model all sources in the vicinity of each candidate and separate the 
fluxes of the candidates from those of their neighbors.
We first use Source Extractor to generate catalogs of sources in the IRAC images and to 
calculate, and subsequently remove, any remaining background pedestal 
present in the mosaic images.
We create IRAC stamps 51 pixels ($30\farcs6$) on a side and centered on 
the position of each candidate in the \textit{HST} F160W imaging.
All sources detected in the F160W images down to a magnitude of $JH_{140}=25$
are added as input to the \textsc{Galfit} models (which accounted for all sources visible in the IRAC images). The input source positions 
are taken from the F160W catalogs, and the initial 3.6$\micron$ magnitude
guess is set to be one magnitude brighter than in $H_{160}$. For the cases 
where the candidate resides close to the edge of the F160W image, we 
supplement the input source list with the positions of sources in the IRAC Source Extractor 
catalog. These additional sources are towards the edges of the IRAC stamp 
and therefore do not affect the model fits to the candidates.
The source positions are constrained to be within $\pm1.5$ pixels 
($\pm0\farcs9$) of the initial guess. There is essentially no constraint 
placed on the source magnitude, though sources with model magnitudes fainter 
than 40 are considered undetected in the IRAC image and are iteratively 
removed from the \textsc{Galfit} model. Finally, we include the sky as a free parameter, though note that the sky values \textsc{Galfit} calculates are all consistent with zero.

Almost all sources are unresolved at 
the IRAC resolution, and so we model them as point sources. 
For this, we created a median point spread function by identifying isolated (no neighbors within 10$\arcsec$), bright ($20 > m_{3.6} >16$) unresolved objects in the IRAC images, identifying such objects in a plane of half-light radius versus magnitude.  We resampled cutouts around these objects by a factor of 10, then centroided, shifted, and median combined them. A total of 81 stars from the five fields were included in the median. This custom-built PSF is used to model all candidates in the IRAC images as well as almost all neighboring sources.
Any sources in the F160W catalog with A\_WORLD greater
than $2\times$ the FWHM of the IRAC PSF are modeled as S\'{e}rsic profiles, though we note that there are very few of these (six across all data considered here). 

Figure~\ref{iracfig} shows the resulting \textsc{Galfit} model fits. For each 
observation, we show the $30\farcs6$ IRAC stamp on top with $12\farcs6$ 
stamps underneath, zooming in on the candidate position. Here, we show 
the IRAC stamp on the left and the residual map(s), with all neighboring 
sources removed, on the right. In five cases, the 3.6\micron\ flux 
at the position of the candidates was too faint to model with \textsc{Galfit}, 
and so the candidates are considered undetected in these IRAC images. 
The purple stars on the residual maps in Figure~\ref{iracfig} indicate cases 
where the candidate was modeled and is left in the residual map for flux 
measurements. 

We measure the 3.6\micron\ flux of the candidates with a circular aperture of 
radius 2.3 pixels ($2\farcs76$ in diameter, or $1.4\times$ the FWHM of the $3.6\mu$m warm mission point response function), shown in purple in Figure~\ref{iracfig}. For consistency, we do this for all sources in the IRAC residual maps, regardless of whether or not the candidate was bright enough to model with \textsc{Galfit}. To estimate the flux uncertainty we randomly place apertures of the same size across the full, background-subtracted IRAC image 
(avoiding flux from real sources using the Source Extractor segmentation map) and fit a Gaussian to the distribution of aperture fluxes.  The 1$\sigma$ flux uncertainty is then the $\sigma$ of this Gaussian fit. Finally, we derive aperture corrections by measuring the flux of our custom PSF in consecutively larger circular apertures and apply a correction of 1.91 to all measured fluxes and flux uncertainties. 

Candidate par2139+0241\_1709 (left-most column in Figure~\ref{iracfig}) has a 
very close neighbor in the F160W image and therefore merits further 
discussion. The candidate and its neighbor are too close together to 
successfully deblend their IRAC fluxes. We therefore consider two limiting 
cases: (1) all IRAC flux belongs to the candidate (middle stamp, 
indicated by a purple star), and (2) all IRAC flux belongs to the neighbor 
(right stamp). In both cases, the best-fit EAZY templates prefer 
high-redshift solutions with $z_{\mathrm{best}} =$ 10.33 and 10.28 for 
cases (1) and (2), respectively. 
We also note that par0953+5153 was observed twice as part of program 14130.
The best-fit \textsc{Galfit} model for one observation of candidate 
par0953+5153\_1655 attributes all IRAC flux to a bright neighbor (top 
right), while the best-fit model for the second observation (bottom right) 
associates some flux with the candidate. 
While we include both flux values in our updated photometric redshift fit, we note that the flux attributed to the candidate is not centered at the 
expected position of the candidate, implying that the flux in fact belongs 
to the neighbor.  

\begin{figure*}

\gridline{\fig{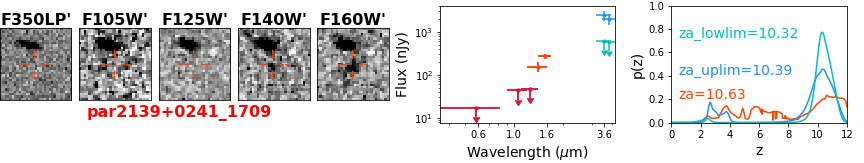}{\textwidth}{}}
\vspace{-14mm}
\gridline{\fig{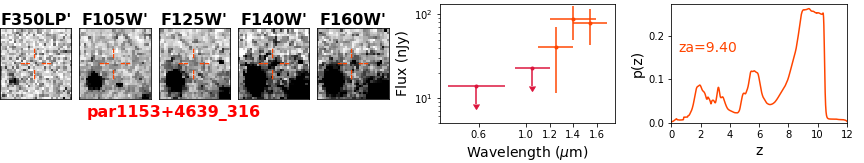}{\textwidth}{}}
\vspace{-14mm}
\gridline{\fig{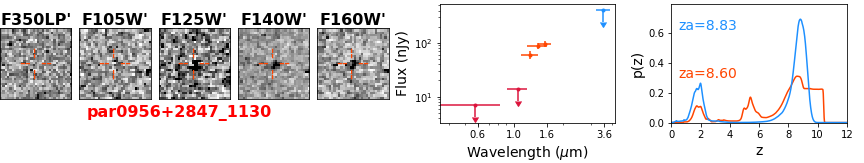}{\textwidth}{}}
\vspace{-9mm}
\caption{Here we show our three $z >$ 8.4 galaxy candidates. {\it Left:} 3$\farcs$0 stamps of each galaxy in the five {\it HST} filters of this survey.  {\it Middle:} The SED of the candidate is presented here with non-detections in the corresponding filter as upper limits. We present the IRAC measurements in blue where available and show their image stamps in Figure \ref{iracfig}. Note that for the case of par2139+0241\_1709 we show both IRAC photometry scenarios, as mentioned in Section \ref{irac}. {\it Right:} The $P(z)$ versus $z$ from EAZY with very clear high probability distributions. The best-fitting redshift is denoted as `za'. The orange distribution is obtained from the EAZY run with only {\it HST} data while the blue comes from adding IRAC constraints where available. }\label{cand1}
\end{figure*}

\begin{figure*}
\gridline{\fig{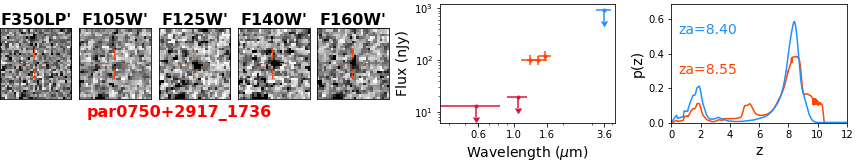}{\textwidth}{}}
\vspace{-14mm}
\gridline{\fig{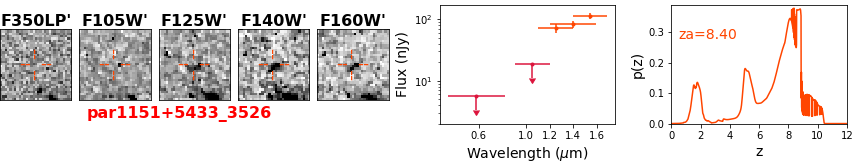}{\textwidth}{}}
\vspace{-14mm}
\gridline{\fig{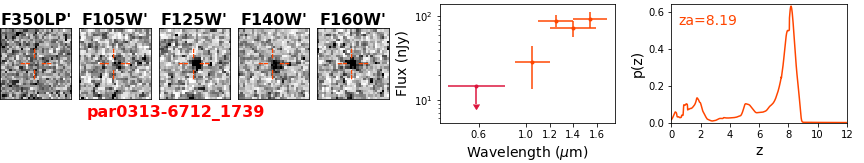}{\textwidth}{}}
\vspace{-14mm}
\gridline{\fig{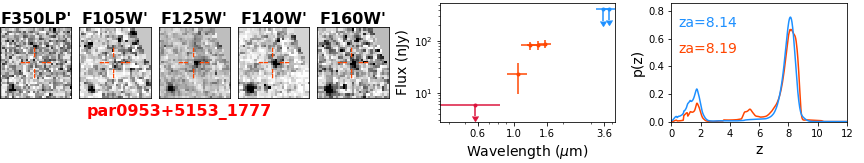}{\textwidth}{}}
\vspace{-14mm}
\gridline{\fig{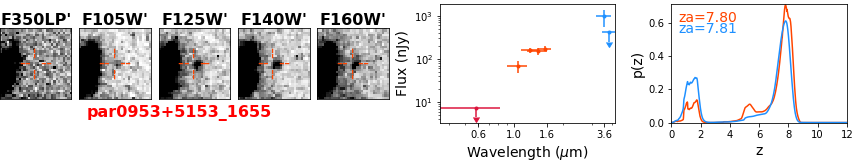}{\textwidth}{}}
\vspace{-14mm}
\gridline{\fig{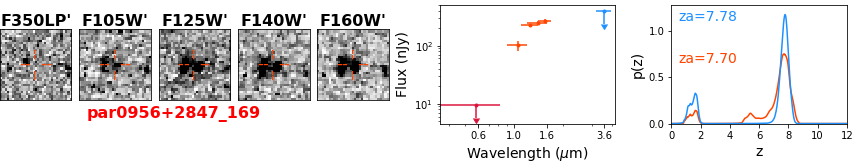}{\textwidth}{}}
\vspace{-14mm}
\gridline{\fig{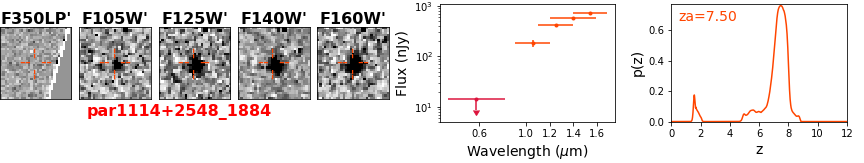}{\textwidth}{}}
\vspace{-14mm}
\gridline{\fig{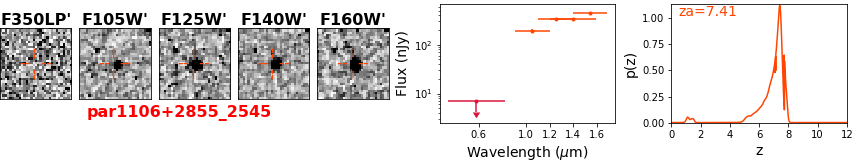}{\textwidth}{}}
\vspace{-10mm}
\caption{Similar to Figure \ref{cand1}, for the seven galaxy candidates at $z <$ 8.4.}\label{cand2}
\end{figure*}

We re-ran EAZY with all measured 3.6$\micron$ fluxes and uncertainties. The resulting redshift probability 
distributions are shown in blue in Figures~\ref{cand1} and \ref{cand2}, and show that all sources but one continue to satisfy our selection criteria.  
For this source, par0807+3606\_25, there is positive IRAC flux, but still at the $<$1$\sigma$ level.  This object has a fairly red $J-H$ color, thus the high-redshift solution expected an IRAC detection, consistent with a red rest-UV continuum.  The IRAC non-detection implies there is a turnover in the SED at $\sim$3 $\mu$m, consistent with the peak of stellar emission at $z <$ 2.  We therefore remove this object from our sample.

For the remaining 11 sources, the addition of the IRAC fluxes often serves to narrow the main, 
high-redshift peak of the EAZY $P(z)$ and, in some cases, also reduces the 
size of any lower-redshift peaks, highlighting the utility of these data.  The fact that only one object was found to be a low-redshift interloper during this step adds confidence to our source selection methodology.
The final catalog of 11 galaxy candidates is described in Table \ref{tab_candidates}.\\


\section{High Redshift Galaxy Candidates}
In this section we present the results of the final set of candidates from this work, including a comparison to previous studies with BoRG[z910] from \citet{Calvi2016}, \citet{Morishita2018}, and \citet{Bridge2019} hereafter referred as C16, M18, and B19, respectively.


\subsection{Galaxy Candidates from this work of BoRG[z910]}
Here we discuss the candidates passing all selection criteria described above, splitting our catalog into two redshift bins based on the best-fitting redshift, of $z =$ 7--8.4 and $z =$ 8.4--10.6. We successfully recover some of the candidates presented in C16, M18, and B19 but not all because of distinct galaxy selection methods, predominantly the different utilization of photometric redshift software.

Due to our sample selection, all of our candidates show a strong peak in their photometric redshift probability distribution function at $z >$ 7.  However, given the relatively small number of filters, each with only moderate depth, other, smaller, peaks are seen.  Most objects show a peak at $z \sim$ 2, which is due to the Balmer/4000\AA\ break.  We also sometimes see a peak at $z =$ 4--6, which we investigated, and found that at these redshifts EAZY prefers a combination of the dust-dominated template and a PEGASE template. While these solutions are possible given our observations, the photometry still clearly prefers a high-redshift solution.

\subsubsection{Candidates at $z =$ 8.4--10.6}
We present our candidates in the higher redshift bin in Figure \ref{cand1}.
Galaxy candidate par2139+0241\_1709 has the highest photometric redshift solution. We obtain two limiting photometric redshifts with the analysis of IRAC data as mentioned in Section \ref{irac}. We take the average of the two photometric redshift solutions and calculate a final value of $z_{best}=$ 10.36, and mag$_{H_{160}} =$ 25.3. This galaxy was previously reported in C16 at $z =$ 10.5. M18 do find this source, but they reject it as being too large, with $r_{0.5} \sim$ 0\farcs5; we measure $r_{0.5} =$ 0\farcs2. M18 use a different reduction method which could lead to the different estimate of half-light radius.
Candidate par1153+4639\_316 has a $z_{best}$ $=$ 9.40 and mag$_{H_{160}} =$ 26.6, and is published here for the first time.  From the stamp images shown in Figure~\ref{cand1} it is clear that the source is in between two very bright sources, and thus magnified due to gravitational lensing. We explore potential lensing magnification for all sources in Section \ref{magnification}.
The last candidate in this redshift bin is par0956+2847\_1130 has mag$_{H_{160}} =$ 26.4 and  $z_{best}$ $=$ 8.83, which is very similar to the redshifts previously found in C16 of $z =$ 8.7, and in M18 of $z =$ 8.8.

\subsubsection{Candidates at $z =$ 7--8.4}
Here we present the galaxy candidates in our lower redshift bin as displayed in Figure \ref{cand2}. All of these galaxy candidates are reported here for the first time.
Candidate par0750+2917\_1736 has mag$_{H_{160}} =$ 26.2 and $z_{best}$ $=$ 8.40. This candidate had been previously observed in \cite{Yan2011} HIPPIES programs 11700 and 11702, but was discarded as a possible high redshift galaxy because it showed flux variability in the $H_{160}$ filter. BoRG[z910] has data from only one epoch for this candidate and therefore we cannot probe variability. However, our analysis indicates that this candidate appears to be resolved as seen in Figure \ref{star_yj}, making it unlikely for this object to be a flaring brown dwarf, which was the explanation by \citet{Yan2011}. Additionally, we note that HIPPIES did not have the $JH_{140}$ filter or IRAC 3.6 $\micron$ which in this BoRG work supports the high redshift solution for this candidate. Galaxy candidate par1151+5433\_3526 is presented at $z_{best}$ $=$ 8.40 and mag$_{H_{160}} =$ 26.3. Candidate par0313-6712\_1739 has a redshift solution at $z_{best}$ $=$ 8.19 and mag$_{H_{160}} =$ 26.5. This galaxy is in one field with a big, bleeding star, however, this candidate is not close to the star and thus we are confident in our photometric measurements for this object. Candidate par0953+5153\_1777 has a clear high redshift solution at $z_{best}$ $=$ 8.14 and has mag$_{H_{160}} =$ 26.5. Candidate par0953+5153\_1655 has $z_{best}$ $=$ 7.81 and mag$_{H_{160}} =$ 25.8, while candidate par0956+2847\_169 has $z_{best} =$ 7.78 with mag$_{H_{160}} =$ 25.3. Both of these sources are also likely magnified by a massive neighbor, we discuss these magnification corrections in Section \ref{magnification}. Galaxy candidate par1114+2548\_1884 has $z_{best}$ $=$ 7.50 and with mag$_{H_{160}} =$ 24.2. Finally, candidate par1106+2855\_2545 has $z_{best}$ $=$ 7.41 and mag$_{H_{160}} =$ 24.8.

\subsection{Comparison with Previous BoRG[z910] Analyses} \label{previous_compare}

We did not recover some of the candidates previously found in C16, M18, and B19. In this section we refer to the sources with the ID style from this work (field name $+$ catalog ID), and discuss the corresponding previous published redshift values. Object par0116+1424\_1365  was presented in C16 at $z =$ 8.4. In our work the source passes all selection criteria previously described and we measure this object to be at $z =$ 7.9. However, in a further follow-up study of some BoRG[z910] fields, \cite{Livermore2018} analyzed imaging in the F098M band that was not available by C16 or in our study. The candidate presented a $3\sigma$ detection in this added filter which supported a low-redshift solution at $z <$ 2; we therefore remove this object from our sample. For the case of object par2228-0945\_777 presented in C16 with $z =$ 8.4 and in M18 with a $z =$ 9.0, our work finds a similar redshift at $z =$ 8.76. However, we remove this candidate because our photometry indicates a S/N$_{350}$ = 2.143, thus just missing the selection criteria we set to S/N$_{350}$ $<$ 2.0 as described in Section \ref{machine_learn}. We inspected the F350LP image at this location, and while there is no obvious signal, there is a signal at this position in F105W (at 2.5$\sigma$ significance), showing that this source is at $z <$ 8. B19 also ultimately discards this candidate as their data preferred a low-redshift solution. 


\begin{deluxetable*}{lccccccccccccc}[ht]
\tablecaption{Final Candidates in Borg[z910]\label{tab_candidates}}
\tablewidth{700pt}
\tabletypesize{\scriptsize}
\tablehead{
\colhead{Field-ID} & \colhead{$\alpha$} & 
\colhead{$\delta$} & \colhead{$H_{160}$} & 
\colhead{S/N$_{350}$} & 
\colhead{S/N$_{105}$} &
\colhead{S/N$_{125}$} &
\colhead{S/N$_{140}$} &
\colhead{S/N$_{160}$} &
\colhead{$r_{0.5}$} &
\colhead{$P(z$=7--11$)$} &
\colhead{z$_{phot}$} &
\colhead{M$_{UV}$} &
\colhead{M$_{UV,corr}$} \\
\colhead{} & \colhead{(deg)} & 
\colhead{(deg)} & \colhead{(AB mag)} & 
\colhead{} & 
\colhead{} & 
\colhead{} & 
\colhead{} & 
\colhead{} & 
\colhead{(pix)} &
\colhead{} & 
\colhead{} & 
\colhead{} &
\colhead{} 
}
\startdata
par2139+0241\_1709 & 324.893847 & 2.675669 & 25.33 & 
-1.3 & -0.1 & 0.1 & 3.9 & 6.8 & 2.2 & 0.91 & 10.36 & -22.21$\pm 0.36$ & -21.77$\pm 0.40$\\
par1153+4639\_316 & 178.449911 & 46.664101 & 26.64 & 
1.3 & 1.6 & 3.3 & 5.5 & 6.4 & 2.0 & 0.64 & 9.40 & -20.74 $\pm 0.70$ & -19.25$\pm 0.91$\\
par0956+2847\_1130 & 149.122759 & 28.792020 & 26.44 &
-0.4 & 0.8 & 5.6 & 7.1 & 6.1 & 1.3 & 0.77 & 8.83 & -20.89$ \pm 0.25$ & ---\\
par0750+2917\_1736 & 117.714116 & 29.271519 & 26.20 &
0.8 & 1.8 & 6.2 & 6.7 & 5.5 & 1.5 & 0.68 & 8.40 & -21.08$\pm 0.30$ & -19.69$\pm 0.76$\\ 
par1151+5433\_3526 & 177.915733 & 54.541262 & 26.27 &
-0.7 & 1.4 & 6.1 & 8.0 & 9.0 & 1.4 & 0.63 & 8.40 & -20.95$\pm 0.23$ & ---\\
par0313-6712\_1739 & 48.410272 & -67.205843 & 26.48 &
0.0 & 2.3 & 7.0 & 5.2 & 5.2 & 1.5 & 0.61 & 8.19 & -20.68$\pm 0.29$ & ---\\
par0953+5153\_1777 & 148.294743 & 51.875234 & 26.54 &
0.3 & 1.7 & 7.0 & 6.9 & 5.8 & 1.3 & 0.65 & 8.14 & -20.68$\pm 0.24$ & -19.98$\pm 0.28$\\
par0953+5153\_1655 & 148.294637 & 51.877122 & 25.80 &
0.5 & 4.7 & 11.4 & 10.4 & 9.4 & 2.0 & 0.65 & 7.81 & -21.34$\pm 0.20$ & -20.32$\pm 1.14$\\
par0956+2847\_169 & 149.113236 & 28.812242 & 25.33 & 0.0 & 5.5 & 17.8 & 17.1 & 14.0 & 1.6 & 0.77 & 7.78 & -21.79$\pm 0.14$ & -21.42$\pm 0.16$\\
par1114+2548\_1884 & 168.657155 & 25.786845 & 24.24 & -0.1 & 10.7 & 25.3 & 37.3 & 36.1 & 1.7 & 0.73 & 7.50 & -22.84$\pm 0.11$ & ---\\
par1106+2855\_2545 & 166.533442 & 28.910642 & 24.77 &
-0.4 & 19.0 & 34.8 & 30.7 & 37.3 & 1.2 & 0.62 & 7.41 & -22.27$\pm 0.16$ & ---\\
\enddata
\tablecomments{This table presents the final catalog of high-redshift galaxy candidates selected in this work from the BoRG[z910] dataset. Column 1 is the field followed by the object ID. Columns 2-3 are the RA and DEC calculated in degrees. Column 4 is the magnitude in the H-band. Columns 5-9 are the calculated signal-to-noise values in the 0$\farcs$4-diameter circular aperture in the {\it HST} bands. Column 10 is the half-light radius of the object in pixels on our 0$\farcs$1 scale. Column 11 presents the integral of the $P(z$=7--11$)$ that we require to be higher than 60$\%$ in our selection criteria. Column 12 presents the photometric redshift with highest probability as calculated with EAZY. Column 13 is the calculated absolute magnitude M$_{UV}$ with uncertainty, and column 14 presents the corrected M$_{UV}$ with uncertainty for candidates presenting magnification from nearby low redshift sources. Note that the z$_{phot}$ and M$_{UV}$ values come from the included IRAC data where available. See Section \ref{irac} for the IRAC analysis.}
\vspace{-9mm} 
\end{deluxetable*}

Candidate par0852+0309\_1677 was reported in C16 with $z =$ 7.6, and in B19 with $z =$ 7.7, in this work we calculated a slightly lower redshift probability at $z =$ 7.29, which placed it initially in our sample, but we ultimately removed it during the color-color analyses shown in Figure \ref{star_yj} for being a likely stellar contaminant.

Next we consider previously published high-redshift candidates which never entered our initial high-redshift catalog. Candidate par0116+1424\_1120 was reported in C16 with $z =$ 7.9, and is also presented in B19 at $z =$ 8.0. We discard this candidate because our data prefer a low-redshift solution. Here we note that B19 had additional data for this source in filters F814W and F098M, which were not available to us at the time of this analysis. C16 candidate par1102+2913\_721 with $z =$ 7.3 also had a secondary redshift peak at lower redshift in C16, and in our work, {\it EAZY} weighted the lower peak higher and thus our results show this source to be more likely at $z \sim$ 1.  Finally, candidate par1151+3402\_517 reported in C16 and in B19 at the same redshift of $z =$ 7.6 has a similar best-fit redshift in our catalog, but more than half of its integrated p(z) is at lower redshifts, and therefore does not pass our selection criteria for $P(z =$ 7--11$)$ $\geqslant$ 0.6. 
Candidate par2134-0707\_651 in C16 is not found in our survey because EAZY cannot constrain the redshift because the S/N  in all bands are found to be less than 4.0.  Additionally, the SED of the galaxy looks very red, and we do not see a significant difference between the different fluxes in the filters that would indicate a Lyman break at any redshift.

From M18, we miss their highest redshift source, in field 2140+0241 and at $z =$ 10.0, as we calculate the p(z) of this source to peak at $z \sim$ 6 with our photometry. We tried an additional EAZY run including the IRAC 3.6$\micron$ photometry analyzed and reported in \cite{Morishita2018} with a mag$_{3.6 \mu m} =$ 23.8, where we find a best redshift solution at an even lower redshift of $z \sim$ 2. In this case we note that our use of {\it EAZY} is different compared to M18 as we do not utilize a prior to calculate the $P(z)$.
\subsection{Surface Density}
In this subsection, we analyze our sample of galaxies to calculate their surface density on the sky, allowing a meaningful comparison to results from other surveys, as well as previous studies with these same BoRG[z910] data.
Over the 90 fields analyzed in this work we calculate a total area of 424 arcmin$^2$ (0.118 degree$^2$).  This area was calculated by summing the pixels in the F160W weight map which had values greater than 100 (e.g., accounting for the area which received photons and was not affected by cosmic rays or bad pixels).

\begin{figure}[t!]
\centering
  \includegraphics[width=0.5\textwidth]{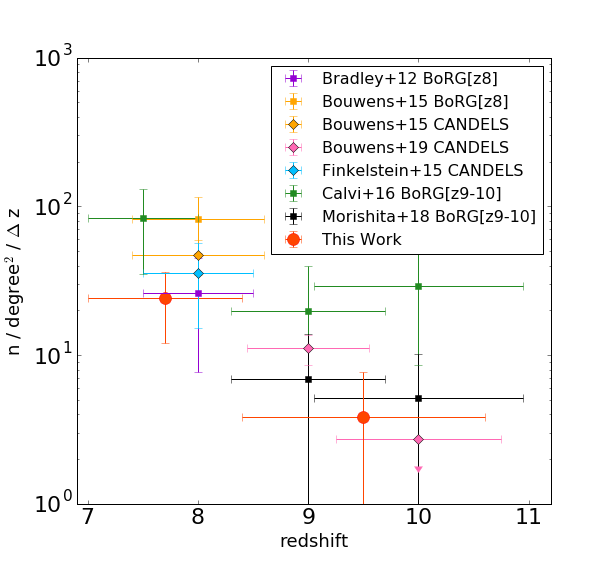}
\caption{Surface density versus redshift of galaxy candidates with magnitude mag$_{H_{160}} < $ 26.0 from this work compared to other surveys. The sources in this work are represented by the orange circles. From the total area analyzed in BoRG[z910] of $\sim$0.12 degrees$^2$, we find four galaxies in the redshift bin of $z =$ 7--8.4 and one galaxy in the redshift bin of $z =$ 8.4--10.6 in this magnitude range. We compare our results with those of surveys in CANDELS, BoRG[z8], and previous studies of BoRG[z910]. The data points presented from other studies are calculated based on the number of galaxies they discovered divided by the area covered in degrees and by the redshift bin width of the respective survey. At lower redshift we are consistent with \citealt{Bradley2012}. At higher redshift we are consistent with the previous studies in BoRG[z910], finding a higher number of candidates compared to \citealt{Bouwens2019} at $z \sim$ 10 from the CANDELS fields.}\label{surf_dens}
\end{figure}

Due to the relatively bright limiting magnitudes of these data compared to other surveys, we calculate the surface density for sources with mag$_{H_{160}} < $ 26.0, and utilize the same magnitude cut for our comparison samples.
Brighter than this magnitude cut, our sample contains four candidates at 7.0 $\leq$ z $\leq$ 8.4 and one galaxy candidate in the redshift bin of  8.4 $<$ z $<$ 10.6.  We calculate the surface density as the number of galaxies per total area observed per unit redshift.  In these two redshift bins, we calculate the surface density (units of \#/deg$^2$/$\Delta z$) as 24.2 $\pm$ 12.1 at $z \sim$ 7.7, and 3.85 $\pm$ 3.85 at $z \sim$ 9.5. These quoted errors assume a Poissonian uncertainty. We calculate the expected additional uncertainty due to cosmic variance with the QUICKCV code \citep[][using the updated version from \citealt{Moster2011}]{newman2002}, adopting a field geometry of 2.2$^{\prime}$ $\times$ 2.12$^{\prime}$ for 90 independent fields (for a total of 424 arcmin$^2$). For this rough estimate, we assumed a bias of 10, extrapolating the results for bright galaxies at $z \sim$ 7 from \citet{Harikane2016} and \citet{Baronenugent2014} to higher redshifts.  This calculation estimates the fractional uncertainty from cosmic variance to be 0.047 and 0.036 at $z \sim$ 7.7 and 9.5, respectively.  Our measurement uncertainties are therefore dominated by Poisson noise. 

These surface densities are compared to those from other studies in Figure~\ref{surf_dens}. As can be seen, our results are consistent with other recent measurements from both BoRG[z8] and BoRG[z910], as well as the CANDELS survey.  Here we discuss those comparison samples in detail, and we remind the reader that we only consider sources brighter than mag$_{H_{160}} =$ 26 in this analysis, and in the comparison samples below.

From pure parallel imaging at $z \sim$ 8, \citet{Bradley2012} found two galaxies over 274 arcmin$^2$ from BoRG[z8], \cite{Calvi2016} found three galaxies at 7 $<$ z $<$ 8 over a 130 arcmin$^2$ from the first release of BoRG[z910] analysis, and \cite{Bouwens2015} found six galaxies over the 218 arcmin$^2$ of highest quality imaging in the BoRG[z8] and HIPPIES datasets.  As shown in Figure~\ref{surf_dens}, our results are highly consistent with those from Bradley et al., but lower than that from Calvi et al.\ and Bouwens et al.\ by $\sim$2$\sigma$.  Comparing to Bouwens, our BoRG[z910] data contains additional imaging in the F140W filter, which should improve the reliability of galaxy selection at $z \geq$ 8.  However, the Calvi et al.\ study used a subset of the data we use here, yet their sample is larger.  As discussed in \S 5.2, some of their candidates were rejected from our sample because they did not pass our selection criteria while others presented a higher redshift probability at low redshifts.  \cite{Finkelstein2015} found three galaxies over 300 arcmin$^2$ of CANDELS GOODS fields, while \cite{Bouwens2015} found a roughly similar surface density at $m <$ 26 with 15 such sources over 959 arcmin$^2$ from all five CANDELS fields.  The CANDELS surface densities are higher than we find (although at $<$2$\sigma$ significance); however, this can easily be explained as those deeper datasets are more complete at $m =$ 26 than the typical field used here, which is confirmed by our completeness simulations (see Figure~\ref{volumes}).

At higher redshift, from BoRG[z910] \cite{Calvi2016} found one candidate galaxy at 8.0 $<$ z $<$ 9.4, and two candidates at 9.6 $<$ z $<$ 11.5 (both at $m <$ 26), both over 130 arcmin$^2$ (here we discarded their low-z contaminant with F098M emission discussed in Section \ref{previous_compare}). This surface density is much higher than what we find because we do not include two of the galaxies presented by \citet{Calvi2016} as discussed in \S 5.2. Their candidate at 8.0 $<$ z $<$ 9.4 corresponds to par2228-0945\_1677, which we rejected since it appears to be stellar in nature. 
Of their two galaxies in the redshift bin 9.6 $<$ z $<$ 11.5, one of them (par2134-0707\_651) is not selected in this work since we did not find a redshift solution at high redshift for this source. From the completed BoRG[z910], \cite{Morishita2018} found one candidate at 8.0 $<$ z $<$ 9.4, and one candidate at 9.6 $<$ z $<$ 11.5 over a 370 arcmin$^2$.  This result is consistent with our identification of two galaxies at $m <$ 26 across this redshift range, though as discussed in \S 5.2, only one of the galaxies is in common.  Finally, over the CANDELS fields, \cite{Bouwens2019} found three bright galaxy candidates with mag$_{H_{160}} < $ 26.0 over $\sim$ 883 arcmin$^2$ in the redshift bins at 8.4 $<$ z $<$ 9.5, and no galaxies in the bin 9.5 $<$ z $<$ 11.\\

\section{Luminosity Function}

\subsection{Magnification Estimates}\label{magnification}
As can be seen in the Figures~\ref{cand1} and \ref{cand2}, several of our candidates have bright galaxies which are nearby and could potentially be magnifying their brightness via gravitational lensing. Given the small number of bright high-redshift galaxies known, these magnifications could potentially bias the shape of the bright end of the luminosity function, thus we attempt to correct for them following the methodology of \citet{mason2015}.  We use the relation between redshift, apparent magnitude and velocity dispersion developed by \citet{mason2015} to estimate the velocity dispersions for galaxies within 10$^{\prime\prime}$ of our high-redshift candidates.  This provides an estimate of the lensing potential of the nearby bright source, which can be used with Equation 4 from \citet{mason2015} to calculate the size of the Einstein radius.  Using the measured separation between our object of interest and the nearby potentially lensing galaxy, this radius can be used to calculate the magnification.

\begin{figure*}[t!]
\centering
  \includegraphics[width=0.8\textwidth]{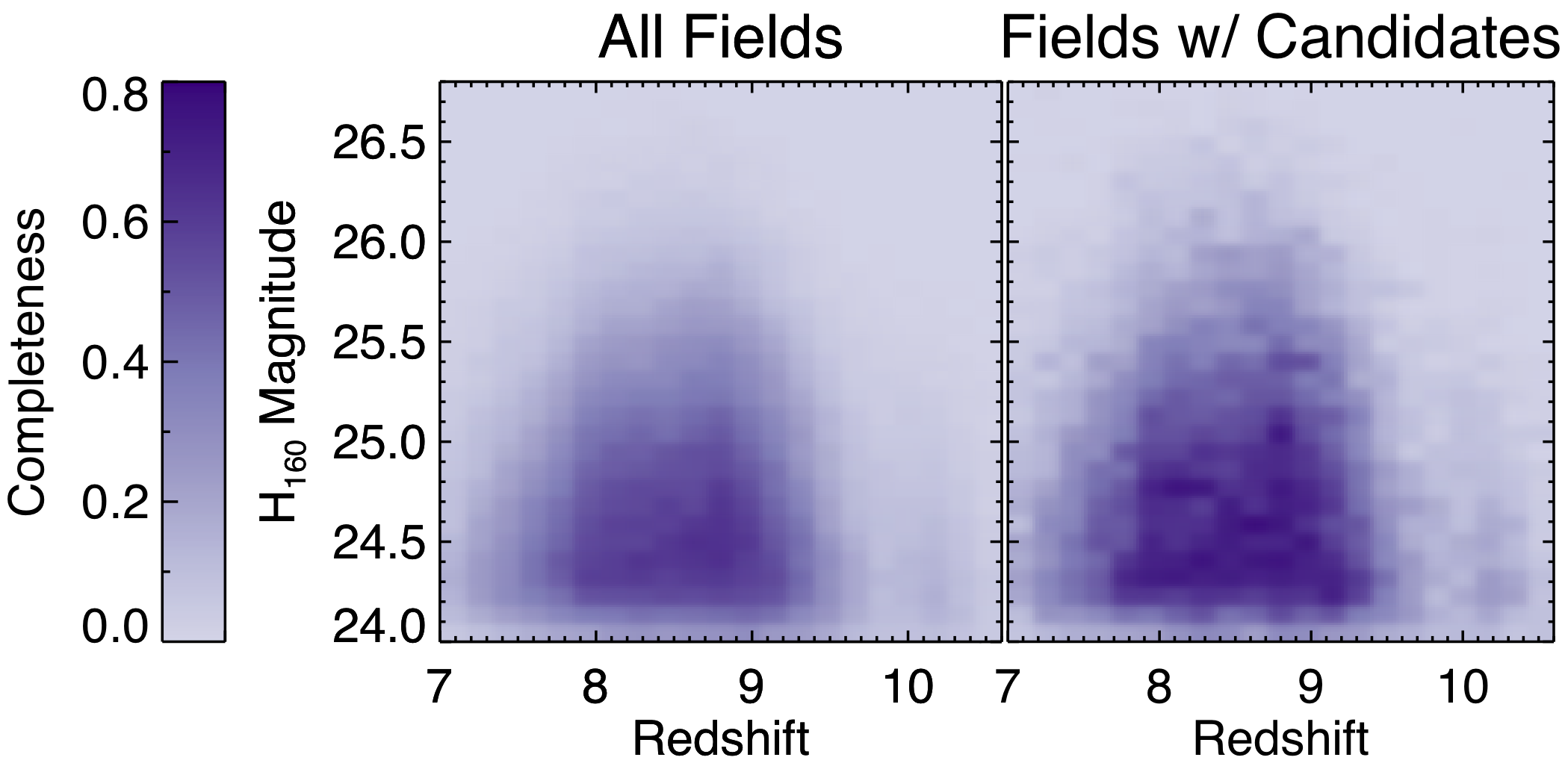}
\caption{The completeness as a function of redshift and apparent magnitude.  These results were obtained via mock source simulations, inserting mock galaxies into images in all 90 fields used in this study.  The left panel shows the average completeness per field, while the right panel shows the completeness in the 9 fields which contain candidate high-redshift galaxies.  The fields where we find our candidates have a higher completeness than the average field.}\label{completeness}
\end{figure*}

One drawback of this method is that it is dependent on our measured photometric redshifts for galaxies in our field. While the dataset we used is sufficient for selecting very high-redshift galaxies, it consists of a single wide optical filter, and several near-IR filters, and thus may not produce reliable photo-z's for galaxies at all redshifts (e.g., galaxies where significant features occur in the optical).  Nonetheless, given the lack of any other data in this field, we use our photometric redshift estimates for these neighbors for this exercise.  

For each potentially magnifying object, we calculate the magnification 1000 times, sampling the photometric redshift probability distribution function for both the high-redshift candidate as well as the potential lens, and perturbing the H-band flux of the lens within its uncertainty.  The fiducial magnification value is the median of these 1000 values, and the uncertainty is the standard deviation.  Out of our total sample of 11 galaxies, we find that six galaxies have a nearby neighbor imparting a significant ($\mu$ $>$ 1.2) lensing magnification, with three objects experiencing $\mu$ $>$ 1.5. 
However, for all three of these the magnifications are very uncertain, often with $\sigma_{\mu} > \mu$.

We included as part of these Monte Carlo simulations the calculation of the absolute rest-UV magnitude. As at $z \sim$ 9.5 the 1500\r{A} rest wavelength is observed near the middle of the $H$-band filter, we calculate the absolute UV magnitude (M$_{UV}$) from the observed $H$-band apparent magnitude and the cosmological distance modulus at the photometric redshift.  In each of these 1000 simulations we  draw a $H$-band flux from the photometric uncertainty, and then correct it for a given simulation's value of magnification (limiting the correction to any magnifying sources with a mean value of $\mu$ $>$ 1.2).  We then convert this flux to an absolute magnitude using the value of the candidate's redshift drawn from the P(z).  This method thus gives us 1000 values of the absolute magnitude for each source, with the uncertainties inclusive of the photometric uncertainty and the lensing uncertainties (when applicable).  In Table \ref{tab_candidates}, we list the median absolute magnitudes (both corrected and un-corrected) for our sources.  We further make use of these 1000 values of the absolute magnitude below when calculating the luminosity function.


\subsection{Effective Volume}
To calculate our luminosity function, we use the effective volume method, where 
\begin{equation}
    V_{eff}(M)=\int \frac{dV}{dz}P(M,z)dz
\end{equation}
where $dV/dz$ is the comoving volume element, and $P(M,z)$ is the probability that an object at a given absolute magnitude and redshift satisfies our sample selection criteria. We estimate $P(M,z)$ using completenesss simulations. We broadly follow the method of \citet{Finkelstein2015}, which we summarize briefly here.

For each of the BoRG[z910] fields in our study, we run a simulation where we place 100 mock galaxies in the images for each field; we run this 200 times, for a total sample of 20,000 mock galaxies per field.  We build SEDs of the mock galaxies to derive their bandpass-averaged fluxes in each of the filters used.  First, we draw a random redshift uniformly over the range 6 $< z <$ 12.  Then for each object, we draw an $H$-band magnitude from a random distribution over 22 $< H <$ 28.  We use a combination of two log-normal distributions, a steep rise from $H =$ 22-24, and a flatter distribution from $H =$ 24-28.  This combination results in $\sim$40\% of the simulated objects having 24 $< H <$ 26, which is the brightness of the majority of our real sources.
We also draw stellar population ages, metallicities and dust attenuation values from log-normal distributions, with typical values of log (age/yr) $\sim$ 7.5, $Z\!\!=$0.2$Z$\sol, and E(B-V)$=$0.15.  The combination of these values produces a UV spectral slope $\beta \approx -$2.0, similar to those observed for bright $z >$ 9 galaxies \citep[e.g.,][]{wilkins2016}.  We use these properties to generate colors from \citet{Bruzual2003} models, normalizing the models to the $H$-band magnitude for a given mock object.  

Mock galaxy images are generated via GALFIT \citep{Peng2002}, with S\'{e}rsic indices drawn from a log-normal distribution tilted towards low values (median $n=$1.8), an axis ratio drawn from a log-normal distribution tilted towards high values (median $b/a=$ 0.75), and a position angle drawn from a uniform random distribution.  We draw galaxy half-light radii using observed relations between galaxy size and their absolute UV magnitudes, using a relation similar to that found by \citet{kawamata2018}, of the form
\begin{equation}
r_{h}(M_{UV}) = 0.94 \times 10^{-0.4 (M_{UV}+21) \beta}~[\mathrm{kpc}]
\end{equation}
where $M_{UV}$ is the absolute UV magnitude of an object, and $\beta$ is the slope of the size-luminosity relation.  We assume $\beta =$ 0.25 for bright ($M < -$ 21) galaxies, and 0.5 for fainter galaxies.  We apply a scatter of 0.2 dex to these sizes to represent the intrinsic scatter at fixed magnitude.  We compared the Source-Extractor measured half-light radii of the recovered sources of the simulations to those measured for our sample of candidate high-redshift galaxies. We found that we could make these comparable, ensuring our simulated population was similar to the real population, but increasing all input simulated radii by 10\%.  This gave a median half-light radii for recovered simulation objects of 1.6 pixels compared to 1.54 pixels for our candidates.

These GALFIT images are normalized to the magnitude for a given object in a given filter, and convolved with the measured point-spread function (PSF).  As our redshift of interest is $z >$ 7, where the galaxy light is detected over only $\sim$1.1--1.6$\mu$m, we use a PSF derived from the F160W image for the simulations.  We chose 10 fields at random, selected stars by finding bright, unresolved and uncrowded objects, and stacked them in each field to make a PSF.  We found that there were no significant variations in the PSF from field-to-field, thus we created a master PSF by stacking the PSFs from these 10 fields, which we use in our simulations.  Finally, the galaxy images are added to a random position of the real image.

\begin{figure*}[t!]
\centering
  \includegraphics[width=0.8\textwidth]{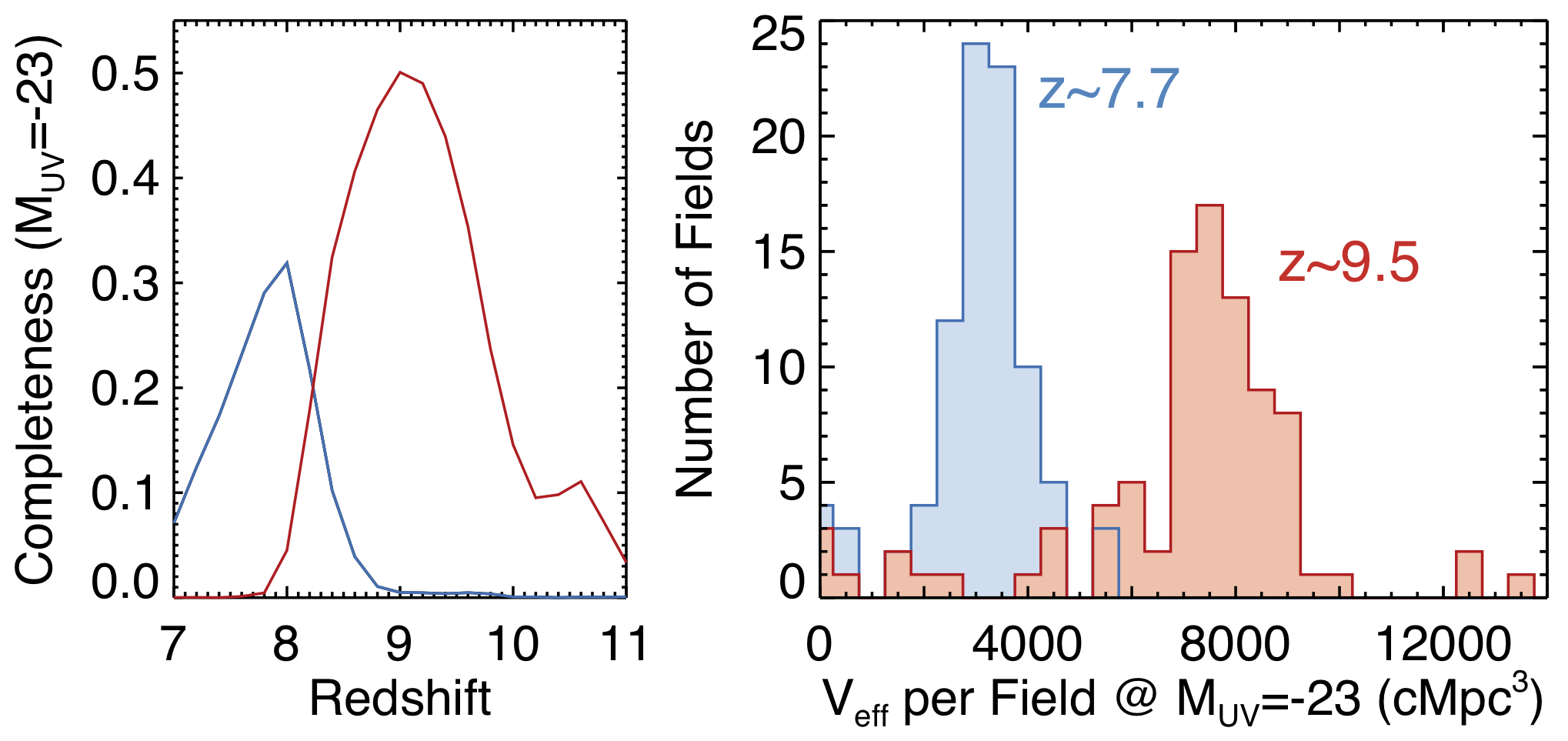}
\caption{{\it Left:} The completeness as a function of redshift for galaxies with $M_{UV} = -$23.  The completeness is lower for galaxies at $z <$ 8 due to the limited photometric information blueward of the Ly$\alpha$ break. {\it Right:}  The distribution of effective volumes for our 90 fields, also calculated at $M_{UV} = -$23.  There is a large spread in the effective volumes, where the fields with very small volumes are highly incomplete due to high stellar densities.
The higher completeness and larger $\Delta z$ for the higher redshift sample results in larger volumes.}\label{volumes}
\end{figure*}

We then measured photometry and photometric redshifts using Source Extractor and EAZY, respectively, in an identical way as done on our real science images, including measuring empirical flux uncertainties based on the aperture sizes and positions of the sources in the images. For each simulation, we matched galaxies in the recovered photometric catalog to the input catalog, counting a source as a recovered match if it was $<$0$\farcs$5 from the input position.  Recovered objects were then subject to the same signal-to-noise and photometric redshift quality criteria as for our real sample.  We calculated the completeness in bins of input magnitude and redshift as the number of fully recovered sources (e.g., found by Source Extractor, and passing all sample selection cuts) divided by the number of input sources per bin.

Figure~\ref{completeness} summarizes these completeness simulations.  The left-hand panel shows the completeness in bins of redshift and $H$-band magnitude, averaged over all 90 fields.  The completeness peaks at $z \sim$ 8--9, which is consistent with the filter set available.  This is also shown in the left-hand panel of Figure~\ref{volumes}, which shows that the completeness for our $z \sim$ 7.7 selection is less than the $z \sim$ 9.5 selection.  
At $z \sim$ 7, only F350LP is fully blueward of the break.  This filter has a very broad response curve, thus the photometric redshift can be quite uncertain, even for bright objects. As the Ly$\alpha$ break shifts through the $Y$-band, the photometric redshift precision increases due to the additional information blueward of the break.  However, at $z \sim$ 9.5, the completeness drops again, as the Lyman break passes through the F140W filter, leaving just the $H$-band free of the break.

In the right panel of Figure~\ref{completeness}, we show a similar figure as the left panel, only here we only include fields which contain our candidate objects.  As may be expected, the average completeness in these fields is somewhat higher, due to a combination of lower stellar densities in these fields, and sometimes deeper photometric depths.  However, even in these fields at bright magnitudes, the completeness peaks at $\sim$60\%.  This low peak in the completeness highlights the difficulty of using these relatively shallow data to search for faint sources, as our objects of interest are close to the 5$\sigma$ depths of the images. We are therefore detecting objects at magnitudes where we are only partially complete. 

We use these completeness results to calculate the effective volume in bins of absolute UV magnitude. We do this separately for our $z \sim$ 7.7 and $z \sim$ 9.5 samples, requiring the best-fit photometric redshift for recovered simulation objects to be less than or equal to (greater than) 8.4 to be placed in our $z \sim$ 7.7 (9.5) sample.  The effective volume is calculated as
\begin{equation}
V_{eff}(M_{1500}) = \int \frac{dV}{dz}~P(M_{1500},z)~dz
\end{equation}
where $\frac{dV}{dz}$ is the comoving volume element.  We show the distribution of volumes for each field in the right panel of Figure~\ref{volumes}, and the total effective volume for our survey is given in Table~\ref{lf_table}.

\begin{deluxetable}{cccc}
\tablecaption{Number densities for luminosity function}
\tablewidth{700pt}
\tablehead{
\colhead{M$_{UV}$ }
& \colhead{Number} & \colhead{$\phi$} & 
\colhead{V$_{eff}$} \\
\colhead{} & \colhead{} & 
\colhead{(10$^{-6}$ Mpc$^{-3}$ mag$^{-1}$)} &
\colhead{10$^4$ Mpc$^3$}
}
\startdata
\multicolumn{4}{c}{$8.4 < z < 11.0$} \\
\hline
-23 & 0 & 1.0175$^{+2.752}_{-0.253}$ & 65.471\\
-22 & 1 & 2.9146$^{+6.390}_{-0.969}$ & 45.811\\
-21 & 1 & 17.932$^{+36.767}_{-7.298}$ & 9.127\\
\hline
\multicolumn{4}{c}{$7.0 < z \leq 8.4$} \\
\hline
-23 & 1 & 5.6904$^{+11.184}_{-2.413}$ & 29.364\\
-22 & 1 & 7.0378$^{+13.413}_{-3.047}$ & 27.604\\
-21 & 3 & 40.326$^{+81.276}_{-18.802}$ & 4.654 \\
\enddata
\tablecomments{These number density values are plotted in Figure \ref{lum_function}.}
\label{lf_table}
\end{deluxetable}

\subsection{Purity of the Sample}
The only way to directly measure the contamination rate in our sample would be to measure a spectroscopic redshift for every source, which is unfeasible.  Here we thus devise a test to estimate whether our sample of candidates is likely to be heavily affected by contamination, which is a significant possibility given the shallowness of BoRG[z910] compared to other surveys used to study the $z >$ 8 universe.
We do this by taking the very deep imaging available from the Hubble Frontier Fields program (HFF; \citealt{Lotz2017}), making use of the deep imaging in the six blank fields taken in parallel to the deep cluster observations. The advantage of this dataset is that each field is similar in area to the BoRG fields, and the available bandpasses cover the same wavelength range. The HFF data include the same four near-infrared bandpasses, while the optical is split into the F435W, F606W, F814W filters (instead of the  F350LP for BoRG[z910]).

After downloading the reduced data from the MAST archive, we ran
Source Extractor on the imaging in all six fields in the same way as we did for the BoRG[z910] data, and then ran EAZY also with the same parameters. The goal of this initial EAZY run using the true depths of the images is to identify any potentially real $z >$ 8 sources (though the analysis of these sources is beyond the scope of this paper). To identify these sources, we applied the selection criteria as we did on the BoRG[z910] dataset.
As one of our selection criteria is S/N$_{350}$  $< 2.0$, we synthesized this value as the average fluxes (and uncertainties) from all three optical filters.  We also accounted for the difference in pixel size between these images (0$\farcs$06) and the BoRG images when applying the half-light radius criterion.
After applying these criteria, we find a total of 24 $z >$ 8 galaxy candidates (plus an additional seven which were discarded after visual inspection) at $H \sim$ 27.

We then simulated BoRG[z910]-level data by increasing the noise of all sources in the catalogs for all six fields.  We did this by randomly selecting flux errors from the input catalogs from five of the BoRG[z910] fields containing candidates presented in this paper. 
We perturbate the HFF flux values by these newly assigned errors, and then run EAZY on this perturbed catalog with the new flux errors to investigate which of these perturbed sources pass our selection criteria.
Across all six fields we find that only one source not previously selected passes all of our selection criteria.  This object's intrinsic photometric redshift was already fairly high at $z_{phot} =$ 6.47, but with the BoRG[z910] noise it is measured at $z_{phot} =$ 10.66 (both the $Y$ and $J$ bands are scattered to low-significance values, increasing the redshift). However, this candidate is right on the edge of a bad pixel region flagged in the rms map (the WFC3/IR ``death star") and therefore would have been discarded during the visual inspection. This test thus shows that in these six fields, we do not find any contaminants.  While we cannot use the results of this test to conclude our contamination rate is zero, it does imply that contaminants do not dominate our sample.

\subsection{Luminosity Function Measurements}
\begin{figure*}[ht!]
\centering
  \includegraphics[width=1\textwidth]{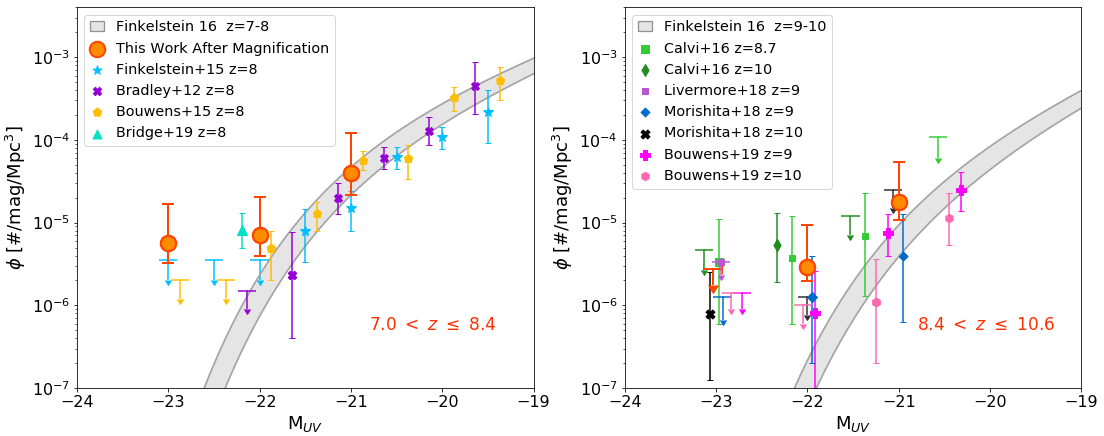}
\caption{Rest-frame UV luminosity functions for our galaxy candidates presented with orange circles in the two redshift bins (7.0 $<$ z $\le$ 8.4), and (8.4 $<$ z $<$ 10.6). For both cases, our data points indicate that the luminosity function may be evolving more slowly at the bright end compared to the faint end. In the lower redshift bin, it is highly consistent at M$_{UV} =-$21 with the luminosity function range at $z =$ 7--8 predicted in \citet{Finkelstein2016}, but it is more discrepant at brighter magnitudes. However, we are very consistent with the averaged LF in \citet{Bridge2019} at M$_{UV} =-$22. In the higher redshift bin, we have more than 1$\sigma$ discrepancy at all magnitudes compared to the predicted luminosity functions at $z =$ 9--10.}\label{lum_function}
\end{figure*}

We now construct the rest frame UV luminosity functions (LF) in the two redshift bins we use to separate our galaxy candidates. The completeness simulations showed that this BoRG[z910] survey is highly incomplete at magnitudes M$_{UV} =-$20 and M$_{UV} =-$19, therefore we do not use the results for these magnitude bins in the remainder of this work. 
The absolute magnitudes used for counting galaxies correspond to those after correcting for magnification as described before. 

The nominal luminosity function is the number of galaxies in a redshift bin per magnitude bin of $\Delta M =$ 1, divided by the volume found after the completion simulations and described in Table \ref{lf_table}.  We follow the methodology of \citet{Finkelstein2015} to calculate both these luminosity function values and their associated uncertainties, using a Markov Chain Monte Carlo (MCMC) method which properly accounts for the Poissonion likelihood of galaxy counting, and returns asymmetric uncertainties. In each step of the chain, for each galaxy in our sample we draw an absolute magnitude from the 1000-step posterior as calculated in \S 6.1.  This ensures that the magnitude uncertainties, which include photometry, photometric redshift, and magnification correction uncertainties, are included in our luminosity function uncertainties (including allowing galaxies to shift magnitude bins.)  We take the median of the posterior (calculated from 10$^5$ steps after a 10$^6$ step burn-in) as our fiducial luminosity function values, and calculate the uncertainty as the central 68\% confidence range of the posterior.
We report these number density values in Table~\ref{lf_table}, and plot them in Figure~\ref{lum_function}.

As we are only calculating the number densities in bright magnitude bins, our study does not possess the dynamic range in luminosity to do a full Schechter function fit.  We thus restrict our analysis to a comparison to previous works.  
We first compare our data to the smoothly evolving luminosity function derived in \citet{Finkelstein2016} from a Schechter fit to various data points from the literature and which has the form
\begin{equation}
\phi(M) = 0.4 \ ln(10) \ \phi^* \ 10^{-0.4 (M-M^*)(\alpha+1)} e^{-10^{(-0.4 (M-M^*))}}
\end{equation}
with parameters
\begin{eqnarray*}
log(\phi^*) = -3.37 - 0.19(z-6)\\
M* = -20.79 + 0.13(z-6)\\
\alpha = -1.91 - 0.11(z-6)\\
\end{eqnarray*}
  
  For the lower redshift bin we plot the predicted luminosity function for $z =$ 7--8, shading in the region bounded by the luminosity function for these two redshifts. We also use the same comparison data points analyzed before to build the surface density; \citep{Finkelstein2015,Bradley2012, Bouwens2015}, in addition to the LF presented in \citet{Bridge2019}. As seen in the left panel of Figure \ref{lum_function}, our data is highly consistent at M$_{UV} =-$21 with previous observations $z =$ 7--8. At M$_{UV} =-$22 we are above the \citet{Finkelstein2016} reference luminosity function, but we do closely agree to the averaged LF at $z =$ 8 produced in \citet{Bridge2019}. We detect one galay in our brightest magnitude bin of M$_{UV} =-$23.  Neither of the papers which studied the CANDELS fields detected a galaxy this bright, but as such bright galaxies are likely highly clustered, this is not surprising.
  
In the right panel of Figure \ref{lum_function} we present the higher redshift bin with the predicted LF shaded from  $z =$ 9--10 along with the comparison data from \citep{Calvi2016, Livermore2018, Morishita2018, Bouwens2019}. Note that in this figure the redshift of comparative data points from previous results is subdivided in galaxies at redshift $z \sim$ 9 and $z \sim$ 10 so that the reader can better observe the distribution of redshifts compared to the luminosity function. In this figure we note that we differ by $>1\sigma$ from the predicted smooth evolution of the luminosity function from \citet{Finkelstein2016} at M$_{UV} =-$21 and $-$22.  We do not find any galaxies in our brightest M$_{UV} =-$23 magnitude bin.  \citet{Morishita2018} has published the only galaxy this bright, which is their $z \sim$ 10 candidate that we did not recover.  At M$_{UV} =-$22, our results are consistent with both redshift bins of C16 as well as the $z =$ 9 result of M18, but are higher than the M18 $z =$ 10 and CANDELS-field results.  Overall, all published results at these high redshifts show significant scatter.  We note that our finding of a higher volume density than previous works is not at odds with our results of a similar surface density in Figure~\ref{surf_dens}.  For the latter, we used the area from all fields, while as shown in Figure~\ref{volumes}, many fields do not contribute significant selection volume, due to lack of depth, high stellar density, or other factors.

\subsection{Implications}
The evolution of the luminosity function can be used as a useful constraint on cosmological simulations \citep[e.g.,][]{yung19,vogelsberger19}.  Our results, along with others based on pure-parallel datasets, have the advantage over contiguous-field surveys such as CANDELS \citep[e.g.,][]{Finkelstein2015,Bouwens2019} in that we better probe the full density field.  As we demonstrated above, we therefore measure number densities with uncertainties dominated by Poisson noise rather than cosmic variance.

Using the inclusive photometric redshift selection technique, our results imply fairly high number densities for bright galaxies -- higher than contiguous-field surveys, and at the high-end of previous pure-parallel survey estimates.  This result can in-part be explained by the fact that our selection does not rely on hard color-cuts, which can exclude otherwise valid candidate high-redshift galaxies.  However, it could also imply that our sample suffers from significant sample contamination.  As described above, we have utilized machine learning to attempt to eliminate contaminants of spurious origin.  We have utilized a combination of colors and source size measurements to remove stellar contaminants, and we have utilized {\it Spitzer}/IRAC imaging (where available) to eliminate low-redshift contaminants. Therefore, to the best that the available data allows, our candidates appear to be valid high-redshift candidates.  

However, we should proceed with caution, as shown in \S 5 several groups have analyzed the same dataset and only have partially overlapping candidate lists.  A clear path forward is spectroscopic confirmation.  This is presently possible with 8-10m class telescopes (if Ly$\alpha$ is viable), or with ALMA, via the very bright [O\,{\sc iii}] 88 $\mu$m emission line.  
Similar pure parallel surveys with {\it JWST} should obtain more robust results, as they will be much more complete at similar magnitudes, and they will also detect similar galaxies in a larger number of filters, yielding higher confidence in their nature (and thus also be less susceptible to large completeness corrections).  
Should {\it JWST} confirm that the abundance of bright $z >$ 7 galaxies is as high as suggested here, it would imply that the bright end of the luminosity function evolves very shallowly, if at all, throughout the epoch of reionization.  This may imply that the physics regulating star-formation evolve with redshift, as recently suggested by \citet{yung19}, who showed that an evolving star-formation law was needed to explain the evolution of the rest-UV luminosity function.  It could also herald the onset of the first large growing super-massive black holes.

\section{Summary}\label{summary}

We use data from the Cycle 22 BoRG[z910] survey to discover galaxies at $z >$ 7, with an emphasis on $z \sim$ 9.  This survey probes $\sim$90 independent lines of sight with {\it HST}'s pure parallel mode, reducing the influence of cosmic variance on studies of distant galaxies. While the total area of this survey is smaller than others, by probing the full range of the distant galaxy density field, we better constrain the volume density of galaxies in our epoch of interest. From our analysis of the BoRG[z910] dataset we find 12 galaxy candidates, among which we utilized {\it Spitzer/IRAC} 3.6 $\micron$ data for six. The additional IRAC fluxes removed one galaxy candidate which, when including IRAC, appeared more likely to be at $z \sim$ 1. 
 Our final catalog of 11 high-redshift galaxy candidates, including nine new discoveries, contributes substantially to constraints at the bright end of the UV luminosity function. These results are complementary to those from contiguous field surveys, such as CANDELS.

We have studied the bright end of the luminosity function at $z \sim$ 7.5 where we are consistent with previous results at M$_{UV} =-$21, but have higher values at brighter magnitudes. However, at M$_{UV} =-$22 we highly agree with the averaged LF at $z =$ 8 from \cite{Bridge2019}. For the LF at $z \sim$ 9.5, we find that our data points are consistent  with those of other surveys in that they suggest that the bright-end of the LF is evolving much slower than would be predicted from contiguous-field surveys alone \citep{Finkelstein2016}.

These galaxy candidates are optimal targets for follow-up observations with ground-based and space telescopes, such as the {\it James Webb Space Telescope}, which will be the leading instrument to continue the search for galaxies at such high-redshifts. 

\acknowledgments
We thank the financial support for developing this work by the NASA Astrophysics and Data Analysis Program through grant award number NNX16AN47G. Additionally, we received support from the \textit{John W.Cox Endowment for the Advanced Studies in Astronomy} distributed by the Astronomy Department; the \textit{FRI Summer Research Fund} from UT College of Natural Sciences. We would also like to thank the UT Office of Undergraduate Research for providing support through the \textit{Student Researcher Award} to participate in conferences where we present this work, and through the \textit{Undergraduate Research Fellowship} for supporting travel to Keck 1 telescope at the W.M. Keck Observatory for follow-up of the galaxy candidates presented in this work.
We would also like to thank Michele Trenti for providing information on the BoRG survey.


\appendix
\section{Table of Fields}
\startlongtable
\begin{deluxetable*}{lccccccccccccccc}
\tablecaption{Information of fields in Borg[z9-10]\label{fields}}
\tablewidth{700pt}
\tabletypesize{\scriptsize}
\tablecolumns{10}
\tablehead{
\colhead{Field Name} & \colhead{$\alpha$} & 
\colhead{$\delta$} & \colhead{E(B-V)} & 
\colhead{F350LP} & 
\colhead{F105W} &
\colhead{F125W} &
\colhead{F140W} &
\colhead{F160W} &
\colhead{Area} \\ 
\colhead{} & \colhead{(deg)} & 
\colhead{(deg)} & \colhead{} & 
\colhead{m$_{lim}$} & 
\colhead{m$_{lim}$} & 
\colhead{m$_{lim}$} & 
\colhead{m$_{lim}$} & 
\colhead{m$_{lim}$} & 
\colhead{arcmin$^2$} 
}
\startdata
par0058-7200 & 14.58 & -72.01 & 0.309 & 26.86 & 26.17 & 26.17 & 26.19 & 26.03 & 4.60 \\
par0110-7248 & 17.67 & -72.80 & 0.084 & 26.97 & 26.31 & 26.47 & 26.51 & 26.32 & 4.60 \\
par0116+1424 & 19.06 & 14.41 & 0.039 & 27.68 & 27.19 & 27.10 & 27.06 & 26.77 & 4.60 \\
par0118-3410 & 19.68 & -34.18 & 0.026 & 27.52 & 26.53 & 26.58 & 26.82 & 26.54 & 4.59 \\
par0132+3035 & 23.10 & 30.59 & 0.047 & 27.20 & 26.61 & 26.57 & 26.66 & 26.40 & 4.60 \\
par0132-7326 & 23.05 & -73.44 & 0.070 & 27.76 & 26.81 & 27.02 & 27.25 & 26.77 & 4.60 \\
par0133+3034 & 23.48 & 30.57 & 0.041 & 23.37 & 22.33 & 22.44 & 22.43 & 22.26 & 4.60 \\
par0133+3040 & 23.43 & 30.68 & 0.039 & 23.15 & 22.19 & 22.16 & 22.02 & 21.87 & 4.60 \\
par0133+3043 & 23.37 & 30.72 & 0.039 & 24.05 & 23.10 & 24.06 & 22.94 & 24.02 & 4.60 \\
par0235-0356 & 38.80 & -3.95 & 0.022 & 28.52 & 27.66 & 27.67 & 27.88 & 27.69 & 6.35 \\
par0313-6712 & 48.43 & -67.20 & 0.036 & 27.79 & 27.72 & 27.81 & 27.71 & 27.45 & 5.07 \\
par0337-0506 & 54.37 & -5.12 & 0.043 & 27.93 & 27.14 & 27.04 & 27.16 & 26.92 & 4.76 \\
par0553-6005 & 88.39 & -60.09 & 0.055 & 27.67 & 26.91 & 27.07 & 27.09 & 26.91 & 4.61 \\
par0750+2917 & 117.71 & 29.28 & 0.041 & 27.77 & 27.33 & 27.24 & 27.28 & 27.03 & 4.60 \\
par0807+3606 & 121.87 & 36.11 & 0.047 & 27.74 & 26.99 & 27.27 & 27.24 & 27.09 & 4.61 \\
par0833+5238 & 128.48 & 52.64 & 0.033 & 27.62 & 27.32 & 27.02 & 27.36 & 27.03 & 4.60 \\
par0850+4239 & 132.72 & 42.66 & 0.023 & 27.72 & 27.12 & 26.93 & 26.98 & 26.78 & 4.61 \\
par0852+0309 & 133.18 & 3.16 & 0.048 & 27.84 & 27.35 & 27.20 & 27.31 & 26.94 & 4.60 \\
par0925+1359 & 141.31 & 14.00 & 0.030 & 27.81 & 27.35 & 27.16 & 27.26 & 27.09 & 4.60 \\
par0925+3438 & 141.32 & 34.65 & 0.019 & 27.98 & 27.18 & 27.22 & 27.28 & 27.03 & 4.60 \\
par0933+5510 & 143.39 & 55.18 & 0.033 & 28.03 & 27.13 & 27.16 & 27.41 & 27.10 & 4.77 \\
par0948+5757 & 147.03 & 57.95 & 0.014 & 27.58 & 26.96 & 27.00 & 27.07 & 26.79 & 4.61 \\
par0949+5759 & 147.34 & 57.99 & 0.013 & 27.70 & 27.39 & 27.03 & 27.28 & 26.98 & 4.60 \\
par0952+5149 & 148.05 & 51.83 & 0.007 & 28.18 & 27.48 & 27.22 & 27.45 & 27.20 & 5.23 \\
par0953+5150 & 148.33 & 51.84 & 0.008 & 28.00 & 27.20 & 27.31 & 27.18 & 27.00 & 4.60 \\
par0953+5153 & 148.32 & 51.89 & 0.009 & 28.41 & 27.58 & 27.63 & 27.53 & 27.36 & 4.95 \\
par0953+5157 & 148.26 & 51.95 & 0.009 & 28.08 & 27.50 & 27.29 & 27.43 & 27.19 & 4.61 \\
par0955+4528 & 148.82 & 45.48 & 0.011 & 27.62 & 27.02 & 26.99 & 27.09 & 26.96 & 4.60 \\
par0956+2847 & 149.10 & 28.80 & 0.017 & 28.02 & 27.22 & 27.54 & 27.40 & 27.15 & 4.76 \\
par1014+5944 & 153.74 & 59.75 & 0.010 & 27.47 & 26.52 & 27.18 & 27.19 & 27.27 & 4.68 \\
par1017+0544 & 154.47 & 5.74 & 0.019 & 27.86 & 27.21 & 27.12 & 27.24 & 26.87 & 4.60 \\
par1017-2052 & 154.35 & -20.87 & 0.042 & 27.49 & 26.52 & 26.64 & 26.63 & 26.40 & 4.73 \\
par1047+1518 & 161.97 & 15.30 & 0.026 & 28.14 & 27.52 & 27.36 & 27.58 & 27.31 & 4.62 \\
par1102+2913 & 165.68 & 29.22 & 0.028 & 28.09 & 27.40 & 27.31 & 27.38 & 27.27 & 4.60 \\
par1103+2812 & 165.97 & 28.21 & 0.032 & 27.86 & 27.20 & 27.00 & 27.24 & 26.91 & 4.61 \\
par1105+2924 & 166.46 & 29.41 & 0.029 & 27.84 & 27.20 & 27.26 & 27.33 & 27.00 & 4.63 \\
par1106+2855 & 166.53 & 28.92 & 0.025 & 28.18 & 27.52 & 27.63 & 27.46 & 27.39 & 4.61 \\
par1106+3508 & 166.53 & 35.14 & 0.018 & 28.07 & 27.37 & 27.37 & 27.46 & 27.06 & 4.61 \\
par1114+2548 & 168.66 & 25.80 & 0.016 & 27.94 & 27.29 & 27.42 & 27.48 & 27.21 & 4.60 \\
par1135+0746 & 173.94 & 7.79 & 0.034 & 27.77 & 27.21 & 27.12 & 27.22 & 26.91 & 4.60 \\
par1141+2640 & 175.46 & 26.67 & 0.019 & 28.19 & 27.40 & 27.44 & 27.56 & 27.17 & 4.76 \\
par1142+2646 & 175.50 & 26.78 & 0.021 & 27.67 & 27.20 & 27.13 & 27.14 & 26.88 & 4.60 \\
par1142+3019 & 175.64 & 30.32 & 0.019 & 27.74 & 27.25 & 27.14 & 27.37 & 27.11 & 4.60 \\
par1142+3020 & 175.62 & 30.34 & 0.020 & 28.13 & 27.56 & 27.41 & 27.45 & 27.09 & 4.61 \\
par1148+2202 & 177.18 & 22.03 & 0.024 & 27.60 & 27.04 & 27.07 & 26.98 & 26.76 & 4.60 \\
par1151+3402 & 177.91 & 34.03 & 0.019 & 27.65 & 26.94 & 26.86 & 26.90 & 26.71 & 4.59 \\
par1151+5433 & 177.94 & 54.56 & 0.010 & 28.25 & 26.96 & 27.49 & 27.54 & 27.38 & 7.49 \\
par1153+4639 & 178.44 & 46.65 & 0.031 & 28.12 & 27.47 & 27.50 & 27.37 & 27.34 & 4.71 \\
par1159+0015 & 179.97 & 0.25 & 0.031 & 27.84 & 27.11 & 27.05 & 27.11 & 27.02 & 4.60 \\
par1209+4543 & 182.36 & 45.72 & 0.014 & 28.33 & 27.36 & 27.42 & 27.60 & 27.51 & 4.61 \\
par1218+3007 & 184.57 & 30.13 & 0.020 & 28.13 & 27.22 & 27.25 & 27.31 & 26.99 & 4.60 \\
par1229+0751 & 187.36 & 7.86 & 0.023 & 27.97 & 27.06 & 26.86 & 26.99 & 26.72 & 4.62 \\
par1258+4127 & 194.66 & 41.47 & 0.014 & 27.97 & 27.43 & 27.51 & 27.55 & 27.22 & 6.36 \\
par1312+1804 & 198.22 & 18.07 & 0.020 & 27.52 & 26.77 & 26.83 & 26.80 & 26.65 & 4.61 \\
par1333+3131 & 203.39 & 31.52 & 0.011 & 28.35 & 27.38 & 27.51 & 27.53 & 27.35 & 4.60 \\
par1409+2622 & 212.41 & 26.38 & 0.016 & 27.95 & 27.24 & 27.25 & 27.41 & 27.08 & 4.61 \\
par1412+0918 & 213.20 & 9.30 & 0.025 & 27.90 & 27.34 & 27.31 & 27.43 & 27.08 & 4.60 \\
par1421+4724 & 215.34 & 47.41 & 0.012 & 27.90 & 26.84 & 27.06 & 27.31 & 26.98 & 4.59 \\
par1431+0259 & 217.86 & 2.99 & 0.028 & 27.49 & 26.69 & 26.63 & 26.76 & 26.36 & 4.61 \\
par1437-0142 & 219.45 & -1.70 & 0.041 & 28.19 & 27.57 & 27.32 & 27.55 & 27.23 & 4.61 \\
par1437-0149 & 219.37 & -1.83 & 0.042 & 27.98 & 27.13 & 27.27 & 27.22 & 27.12 & 4.61 \\
par1442-0211 & 220.54 & -2.20 & 0.051 & 27.72 & 27.20 & 27.03 & 27.14 & 26.81 & 4.59 \\
par1503+3644 & 225.81 & 36.74 & 0.014 & 27.84 & 27.23 & 26.94 & 27.21 & 26.97 & 4.59 \\
par1519-0745 & 229.77 & -7.77 & 0.095 & 27.78 & 27.17 & 27.05 & 27.20 & 26.96 & 4.60 \\
par1520-2501 & 230.08 & -25.02 & 0.158 & 27.69 & 26.92 & 26.85 & 26.86 & 26.57 & 4.59 \\
par1524+0955 & 231.17 & 9.92 & 0.038 & 27.38 & 27.14 & 26.99 & 27.06 & 26.72 & 4.60 \\
par1524+0956 & 231.02 & 9.94 & 0.040 & 27.97 & 27.43 & 27.24 & 27.40 & 27.08 & 4.61 \\
par1524+0959 & 231.19 & 10.00 & 0.037 & 28.12 & 27.22 & 27.08 & 27.22 & 27.24 & 4.60 \\
par1536+1410 & 234.10 & 14.17 & 0.045 & 28.09 & 27.37 & 27.50 & 27.43 & 27.17 & 4.60 \\
par1558+0811 & 239.57 & 8.20 & 0.037 & 27.96 & 27.09 & 26.89 & 27.16 & 26.95 & 4.60 \\
par1606+1332 & 241.70 & 13.54 & 0.035 & 28.06 & 27.35 & 27.55 & 27.35 & 27.17 & 4.60 \\
par1614+4856 & 243.51 & 48.94 & 0.013 & 28.02 & 27.53 & 27.40 & 27.43 & 27.22 & 5.24 \\
par1619+2540 & 244.83 & 25.68 & 0.046 & 27.90 & 27.26 & 27.46 & 27.36 & 27.13 & 4.60 \\
par1631+3736 & 247.89 & 37.61 & 0.009 & 28.13 & 27.26 & 27.36 & 27.17 & 27.19 & 5.28 \\
par1659+3731 & 254.80 & 37.53 & 0.017 & 27.87 & 27.06 & 27.43 & 27.33 & 26.95 & 4.61 \\
par1708+4237 & 257.11 & 42.62 & 0.023 & 28.28 & 27.32 & 27.20 & 27.43 & 27.16 & 4.61 \\
par1715+0454 & 258.75 & 4.92 & 0.114 & 27.75 & 27.14 & 26.97 & 27.00 & 26.82 & 4.60 \\
par1715+0502 & 258.79 & 5.03 & 0.126 & 27.92 & 27.26 & 27.19 & 27.18 & 27.00 & 4.61 \\
par1737+1839 & 264.41 & 18.65 & 0.058 & 27.69 & 26.94 & 26.98 & 26.98 & 26.75 & 4.60 \\
par1920-4531 & 290.10 & -45.52 & 0.083 & 27.64 & 26.95 & 26.86 & 26.99 & 26.76 & 4.61 \\
par2007-6610 & 301.98 & -66.17 & 0.068 & 27.83 & 26.64 & 26.95 & 27.06 & 26.74 & 4.59 \\
par2057-1422 & 314.34 & -14.38 & 0.048 & 27.78 & 27.18 & 27.10 & 27.01 & 26.81 & 4.60 \\
par2134-0707 & 323.54 & -7.13 & 0.031 & 27.91 & 27.08 & 27.08 & 27.26 & 27.00 & 4.70 \\
par2139+0241 & 324.88 & 2.69 & 0.085 & 28.11 & 27.04 & 26.98 & 27.13 & 27.08 & 4.59 \\
par2140-2309 & 325.15 & -23.17 & 0.047 & 26.95 & 26.35 & 26.18 & 26.33 & 26.01 & 4.60 \\
par2228-0945 & 337.19 & -9.75 & 0.048 & 27.67 & 26.97 & 27.02 & 27.02 & 26.91 & 4.60 \\
par2228-0955 & 337.11 & -9.92 & 0.050 & 27.97 & 27.18 & 27.09 & 27.15 & 26.86 & 4.60 \\
par2253-1411 & 343.37 & -14.19 & 0.042 & 27.98 & 27.30 & 27.30 & 27.29 & 27.03 & 4.60 \\
par2311-1423 & 347.93 & -14.39 & 0.033 & 27.99 & 27.37 & 27.28 & 27.18 & 27.02 & 4.59 \\
par2322-0059 & 350.71 & -0.98 & 0.042 & 27.73 & 26.83 & 26.90 & 26.87 & 26.69 & 4.60 \\
\enddata
\tablecomments{A table with all the information of the fields comprised in the BoRG[z910] survey and analyzed in this work. Column 1 is the field name composed of the coordinates of the center in the field. Column 2-3 are the $\alpha$ and $\delta$ in degrees. Column 4 is the E(B-V) galactic extinction from Schlafly \& Finkbeiner (2011), obtained with IRSA. Columns 5-9 are the 5$\sigma$ AB limiting magnitudes per band calculated within a 0.4'' aperture. Column 10 is the area in arcmin$^2$ covered per field. }
\end{deluxetable*}

\end{document}